\documentclass[amsmath, amsonfts,twocolumn]{revtex4}

\DeclareMathAlphabet{\mathpzc}{OT1}{pzc}{m}{it}

\usepackage[colorlinks]{hyperref}

\usepackage{graphicx}
\usepackage{amssymb,mathrsfs}
\usepackage{wasysym}
\usepackage{stmaryrd}
 
\usepackage{lineno,hyperref}
\usepackage{amsmath,bm}
\usepackage{xfrac}

\newcommand{\xgp}{x_\text{\tiny J}}
\newcommand{\phineu}{\psi}
\newcommand{\Phis}{\psi_\text{\tiny s}}
\newcommand{\Phia}{\psi_\text{\tiny a}}

\newcommand{\Y}{{\mathsf y}}

\newcommand{\cB}{{\cal B}}
\newcommand{\cBs}{{\cal B}_\text{\tiny s}}
\newcommand{\cBa}{{\cal B}_\text{\tiny a}}
\newcommand{\cBac}{\hat {\cal B}_\text{\tiny a}}

\newcommand{\Aa}{A_\text{\tiny a}}
\newcommand{\As}{A_\text{\tiny s}}

\newcommand{\ps}{p_\text{\tiny s}}
\newcommand{\pa}{p_\text{\tiny a}}
\newcommand{\pas}{p_\text{\tiny as}}
\newcommand{\zs}{z_\text{\tiny s}}
\newcommand{\za}{z_\text{\tiny a}}

\newcommand{\cAs}{{\cal A}_\text{\tiny s}}
\newcommand{\cA}{{\cal A}}

\newcommand{\phics}{\phi_\text{\tiny s}}
\newcommand{\phica}{\phi_\text{\tiny a}}

\newcommand{\phis}{\phig_\text{\tiny s}}
\newcommand{\phia}{\phig_\text{\tiny a}}

\newcommand{\cZs}{{\cal Z}_\text{\tiny s}}
\newcommand{\cZa}{{\cal Z}_\text{\tiny a}}

\newcommand{\ys}{{\mathrm y}_\text{\tiny s}}
\newcommand{\ya}{{\mathrm y}_\text{\tiny a}}

\newcommand{\Zs}{Z_\text{\tiny s}}
\newcommand{\Za}{Z_\text{\tiny a}}

\newcommand{\zss}{Y_\text{\tiny s}}
\newcommand{\zaa}{Y_\text{\tiny a}}

\newcommand{\ta}{\text{t}}

\newcommand{\Gs}{G_\text{\tiny s}}
\newcommand{\Ga}{G_\text{\tiny a}}

\newcommand{\als}{\alpha_\text{\tiny s}}
\newcommand{\ala}{\alpha_\text{\tiny a}}
\newcommand{\bes}{\beta_\text{\tiny s}}
\newcommand{\bea}{\beta_\text{\tiny a}}

\newcommand{\beff}{b_\text{\tiny eff}}

\newcommand{\ksz}{k_0}
\newcommand{\ks}{k_\text{\tiny s}}

\newcommand{\sphi}{[\phig]}
\newcommand{\sphip}{[\phig']}
\newcommand{\mphi}{\overline{\varphi}}
\newcommand{\mphip}{\overline{\varphi'}}
\newcommand{\bad}{b_\text{\tiny add}}

\newcommand{\ci}{\mathrm{i}}

\newcommand{\phig}{\varphi}

\newcommand{\X}{{\mathsf x}}

\renewcommand{\th}{\theta}

\newcommand{\grad}{\nabla}

\newcommand{\ex}{{\bm e}_x}
\newcommand{\ey}{{\bm e}_y}

\newcommand{\toutin}{\left\{\begin{array}{l}}
\newcommand{\toutind}{\left\{\begin{array}{ll}}
\newcommand{\toutint}{\left\{\begin{array}{lll}}
\newcommand{\toutout}{\end{array}\right.}

\newcommand{\dsp}{\displaystyle}
\def\beq{\begin{equation}}
\def\eeq{\end{equation}}

\begin{document}

\title{Modelling Autler-Townes splitting and acoustically induced transparency  
in a waveguide loaded with resonant channels   }
\author{Richard Porter }
\address{
School of Mathematics, University Walk, University of Bristol, Bristol, BS8 1TW, United Kingdom
}
\author{Kim Pham }
\address{IMSIA, ENSTA Paris - CNRS - EDF - CEA, Universit\'e Paris-Saclay, 828 Bd des Mar\'echaux, 91732 Palaiseau, France
}
\author{Agn\`es Maurel}
\address{Institut Langevin,  ESPCI Paris, Universit\'e PSL, CNRS, 1 rue Jussieu, 75005 Paris, France
}

\begin{abstract} 
We study  acoustic wave propagation in a waveguide loaded with two resonant side-branch  channels. In the low frequency regime, one-dimensional models are derived 
in which the effect of the channels are reduced to jump conditions   across the junction. When the separation distance  is  on the scale of the wavelength,  which is the case that is usually considered, the jump
 conditions involve a single channel and acoustically induced transparency  (AIT) 
 occurs due to out-of-phase interferences between the two junctions. 
 In contrast, when the separation distance is subwavelength, a single junction  has to be considered and the jump conditions  account for the evanescent field coupling the two channels. 
 Such  channel pairs can scatter as a dipole resulting in perfect transmission 
 due to Autler-Townes splitting (ATS). We show that combining the two  mechanisms offers  additional degrees of freedom  to control  the transmission spectra. 
\end{abstract}

\maketitle

\section{Introduction}
 
  \begin{figure}[b!]
\centering
\includegraphics[width=1\columnwidth]{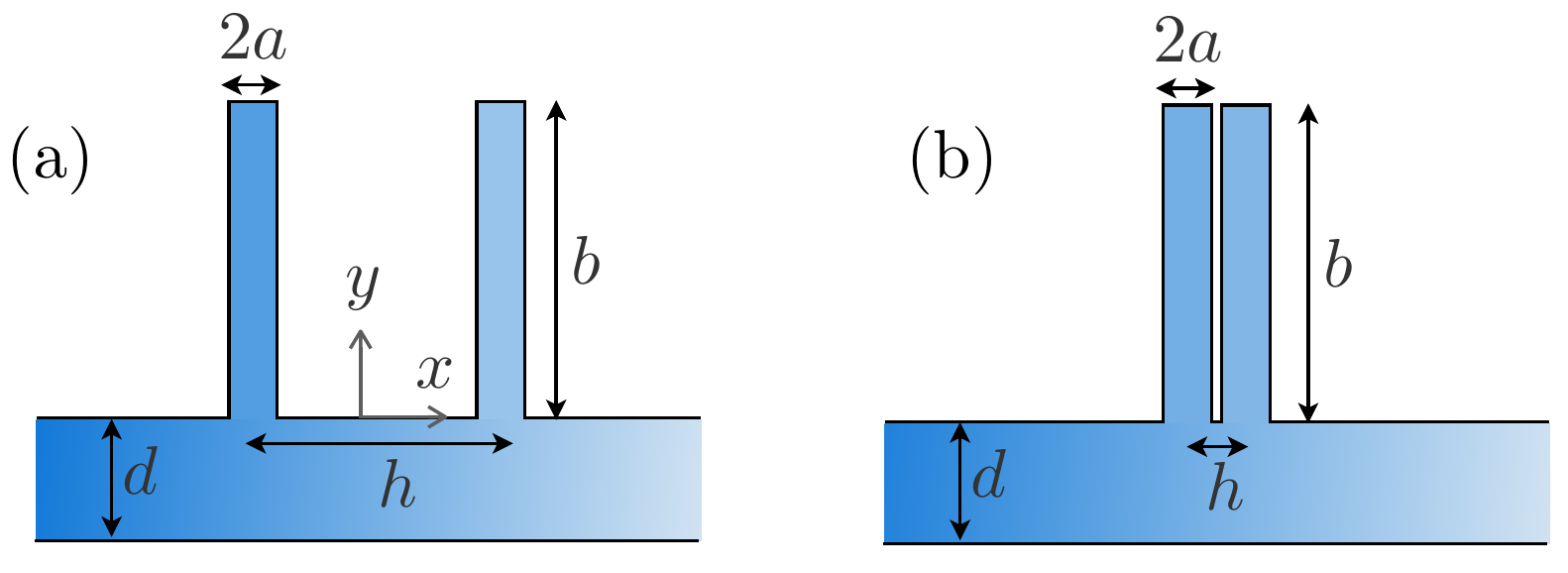}
	\caption{ Propagation in a narrow guide loaded by two channels with
	 separation distance of the order of the wavelength  (a) or of subwavelength scale (b).}
	\label{Fig1}
\end{figure}
Wave propagation in  narrow guides  is a practical exam\-ple of one-dimensional propagation. In this context, re\-so\-nant scatterers, located inside the guide or attached to its walls, can  strongly modify the propagation of the guided waves. 
When the scatterers are distributed periodically along the guide with spacing about half-wavelength, re\-so\-nant Bragg interference take place which can be tuned close to the local scatterer resonance 
\cite{bagwell93,kushwaha1998,el2008,kaina2013,theocharis2014,santillan2014,merkel2015,leclaire2015}. The appearance, within the Bragg band-gap,  of a propaga\-ting branch 
 resulting in acoustically  induced transparency (AIT)  has been studied in detail and the associated Fano resonance has been characterized, see {\em e.g.} \cite{el2008}. 
More recently, directly or strongly coupled resonators have been shown able to produce similar transparency window  
  attributed to the classical analog of Autler-Townes splitting (ATS) \cite{anisimov2011,peng2014,jin2018,liu2019,cheng2019,wang2021}.    AIT is due to destructive  interference  of the  waves  propagating back and forth between the two scatterers.   In contrast ATS does not involve propaga\-ting waves as the scatterers are directly coupled via the evanescent field.  
\vspace{.2cm}

We consider two  narrow channels with closed ends supporting quarter-wavelength resonance attached to one wall of the main waveguide  (figure \ref{Fig1}). When the distance between the channels is of the order of the  wavelength (or greater), the propagation in the guide is well described by a one-dimensional  model in which  the effect of each channel is encoded in    jump conditions
across the channel junction.  Accounting  for the effect of the evanescent field in the junction region improves the accuracy of the model by modifying the form of the jump conditions \cite{wang2014,wang2018,cervenka2018}. So far, this approach has not been adapted to the case where the two channels are so close that they can couple directly through the evanescent field. 
We provide such a model using asymptotic analysis in the low frequency regime. To stress the difference between the two situations, we first revisit the classical case for large separation distances between the channel junctions. We show that the   jump conditions are obtained by considering 
a locally-incompressible region governed by Laplace's equation. This problem is solved exactly  by exploiting a conformal   mapping which provides an explicit expression of the added length in two-dimensions in excellent agreement with previous estimates \cite{cervenka2018,cervenka2019}.
We then extend this analysis to the case where the junction  includes two channels with subwavelength separation distance.

\vspace{.2cm} 
 
The paper is organized as follows. In \S \ref{S2}  the derivation of the two models is presented resulting in the jump conditions \eqref{saut1} and \eqref{saut2}. The scattering  by a  single channel is presented in \S \ref{S3} and 
 this provides an important refe\-ren\-ce to our discussion relating to the scattering by a  set of two channels, see \S \ref{S4}. We can already note that for large separation distance $h$ (termed model a) neglecting  the  effect of the evanescent field in the transition conditions provides reasonable predictions since the resonant  mechanism 
  of AIT is captured \cite{el2008,chesnel2018}. In contrast, the strong evanescent field excited in the junction region  of  two close channels (termed model b) is  the basis of the dipolar resonance involved in the ATS. Finally, we envisage in \S \ref{S5} a situation involving both resonances using scatterers comprised of a pair of close channels. 
 The Appendices contain additional modelling information, including the derivation of the parameters entering  in the model a and which are found using a conformal mapping. In the study, the results of the effective models are compared with direct numerics based on a multimodal method similar to that used in \cite{maurel2019}.

\section{The effective models}
\label{S2}
\begin{figure}[h!]
\centering
\includegraphics[width=1\columnwidth]{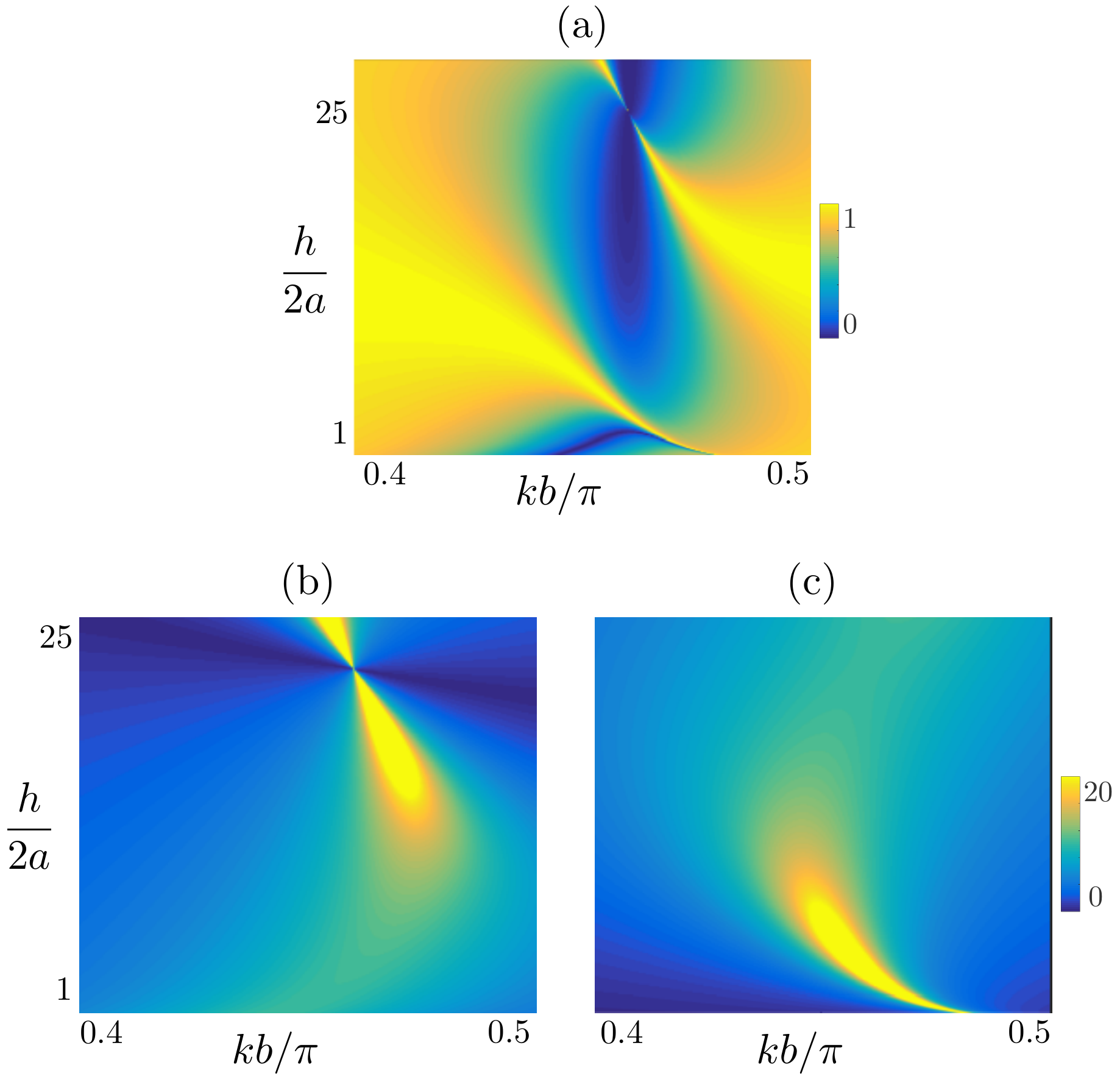}
	\caption{(a) Transmission in the main guide against non-dimensional frequency 
	$kb/\pi$ and  separation distance  $h/2a$
	($b=d=20a$). (b) Sum and (c) difference of the field amplitudes at the top of the channels.
	 }
	\label{Fig2}
\end{figure}

In this section, we provide the derivation of two diffe\-rent models whose range of validity depends on the se\-pa\-ra\-tion distance, $h$, between the two channels.
To set the scene, we aim to model  the perfect transmission illustrated in figure \ref{Fig2}(a)  for $h/2a\sim 1$ and $h/2a\sim 25$ in the reported example. 
We  notice that perfect transmission is associated with large field amplitudes 
 within the channels being either in-phase for large $h$ (panel (b)) or out-of-phase for small $h$ (panel (c)). 

\begin{figure}[t!]
\centering
\includegraphics[width=.9\columnwidth]{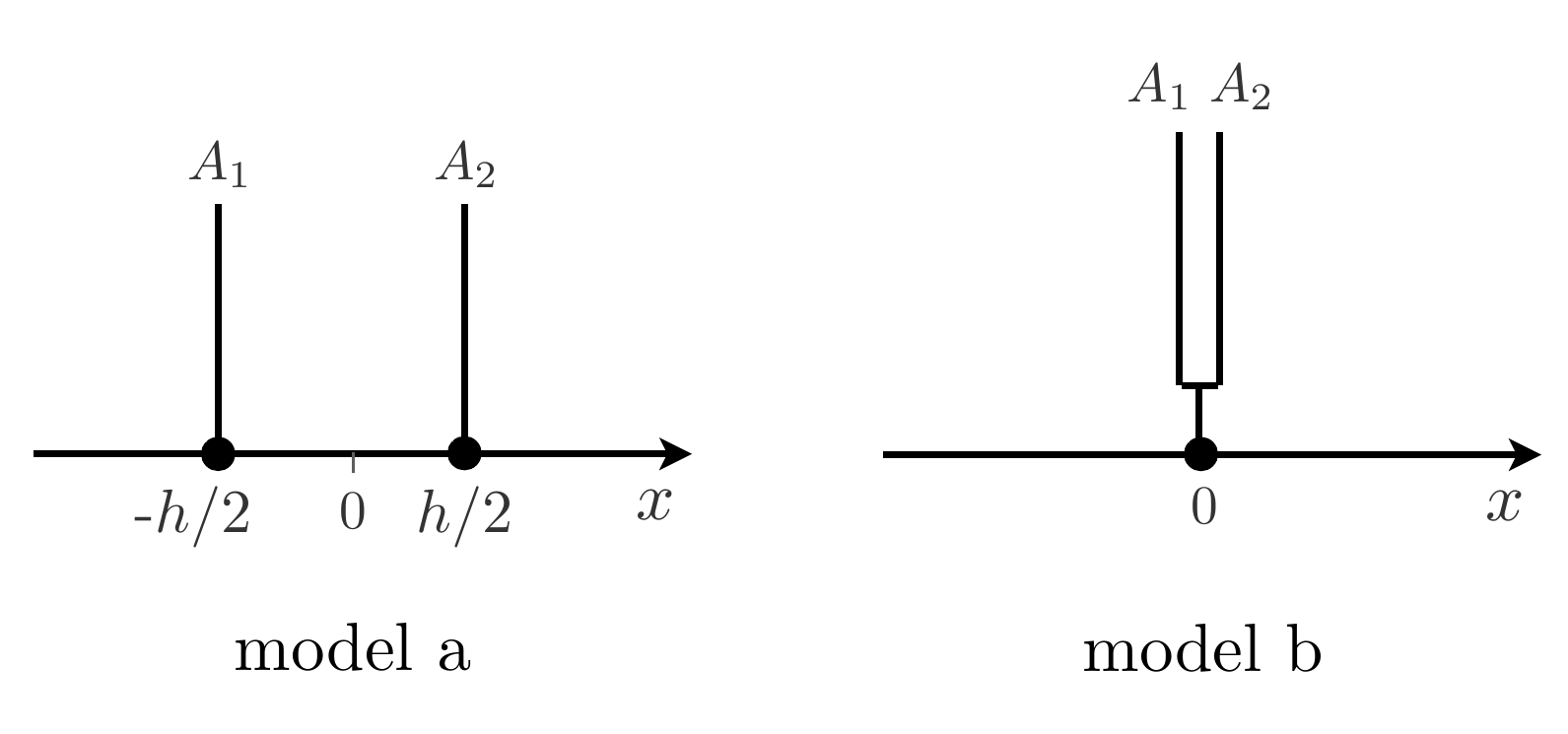}
	\caption{The two one-dimensional models :  model a for  $kh = O(1)$  resulting in 
	jump conditions \eqref{saut1} at two junctions $\xgp=\pm h/2$, and model b for 
	 $kh\ll 1$ resulting in jump conditions \eqref{saut2} at a single junction $\xgp=0$.}
	\label{Fig3}
\end{figure}

The jump conditions are obtained using matched   asymptotic expansions which requires general solutions in the guide and in the side channels which capture the dominant far-field behaviour of the narrow side channel opening. The approximation to the solution in the guide and in the side channel far away from the channel ope\-ning (the junction) is elementary. The most challenging part is the inner region in the vicinity of the junction region. 
The two models differ since in the model a the junction includes the effect of a single channel while in the model b the junction includes the effect of the pair of side channels, see figure \ref{Fig3}.

\subsection{Solution far from the junction}
The problem is governed by the Helmholtz equation  $\Delta \Phi+k^2\Phi=0$ for $\Phi(x,y)$.
 In the main channel far away from the opening at $x=0$,  we shall only need that the solution can be approximated by $\Phi(x,y)\simeq \phig(x)$ governed by $\phig''(x)+k^2\phig(x)=0$. We note that 
\begin{equation}
  \phig(x) \underset{kx\to k\xgp}\sim \phig(\xgp^{\pm}) + (x-\xgp) \phig'(\xgp^{\pm}).
 \label{eq:3.5}
\end{equation}
Similarly, in the $n$th side-channel, $n=1,2$, far away from the opening, we have  $\Phi(x,y) \approx \phi_n(y)$ governed by
$ \phi''_n(y) + k^2 \phi_n (y) = 0$, and we note that
\begin{equation}
 \phi_n(y) \underset{ky\to 0}{\sim} \phi_n(0) + y  \phi_n'(0).
 \label{eq:3.10}
\end{equation}

\subsection{Derivation of the jump conditions}

We define 
$$\mphi=\frac{1}{2}\left(\phig^++\phig^-\right), \qquad 
\sphi=\left(\phig^+-\phig^-\right)$$
to be   the average and the jump of $\phig$ across the junction at $x=\xgp$ ($\phig^\pm=\phig(\xgp^\pm)$). 
Also, we denote $\th=a/d$.  
  
\subsubsection{Solution in the junction region with a single channel}
\label{single}
In the vicinity of the junction between the guide and the channel, we rescale variables on   $d$ with 
  $$x-\xgp=d\X, \qquad y=d\Y.$$
  Accordingly  
 we set $\Phi(x,y) \simeq \Psi(\X,\Y)$ which, 
under the working assumption that $kd \ll 1$, 
approximates the Helmholtz equation as the Laplace equation
\beq\label{Laplace}\nabla^2 \Psi = 0.\eeq
We shall now use that the velocity $\grad\Psi$ at the local scale has to match the velocities in the guide when $kx\to k\xgp^\pm$ and in each channel when $ky\to 0$.  Specifically, we must have
\beq\label{bc1}\toutin
\grad\Psi\underset{\X\to\pm\infty}{\sim} d\phig'(\xgp^\pm)\,\ex=d\left(\mphip\pm\frac{1}{2}\sphip\right)\ex,\vspace{.2cm}\\
\grad\Psi\underset{\Y\to+\infty}{\sim} d\phi_1'(0)\,\ey=-\frac{\ey}{2\th}\,d\sphip,
\toutout\eeq
where $\X\to\pm\infty$ means $\X\to \pm\infty$ in the guide and $\Y\to+\infty$ means $\Y\to +\infty$ in the channel. In \eqref{bc1}, we have used that
$\sphip+2\th\phi_1'(0)=0$. This relation
 corresponds to the continuity of the flux which arises from
integrating \eqref{Laplace} and  
applying the divergence theorem to $\grad \Psi$. 

\vspace{.3cm}

\begin{figure}[h!]
\centering
\includegraphics[width=1\columnwidth]{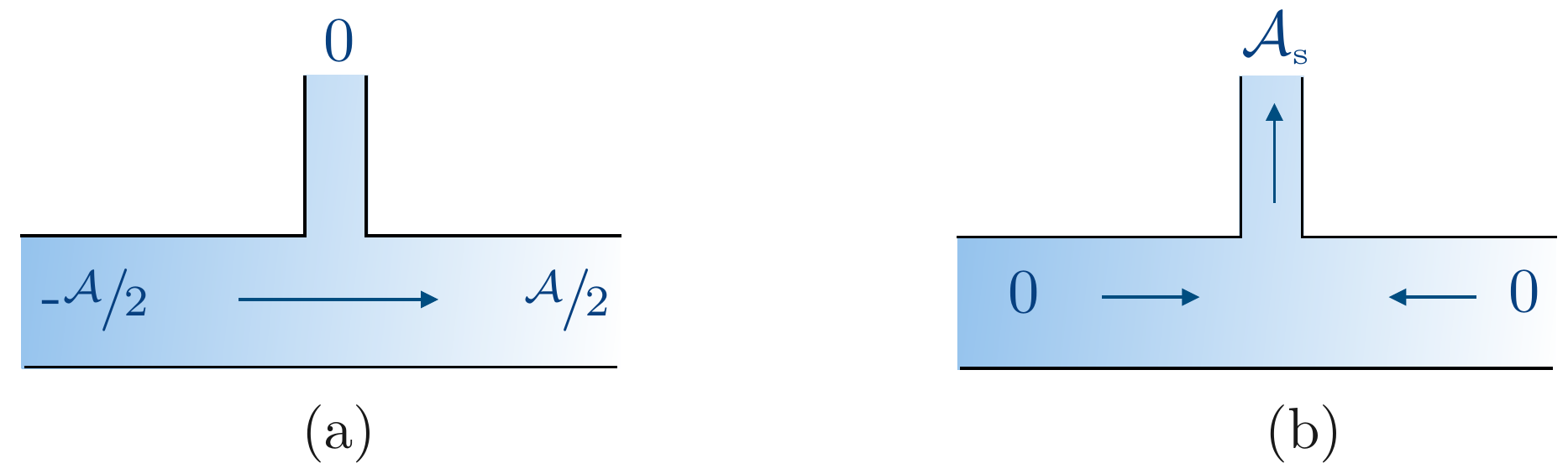}
	\caption{ Elementary problems of potential flows  with \eqref{ag1} set on (a) $\phineu$ and (b) $\Phis$. }
	\label{Fig4}
\end{figure}

The system (\ref{Laplace}-\ref{bc1}) is linear with respect to $\mphip$ and $\sphip$, from which we deduce that $\Psi$ can be sought of the form 
\beq\label{psi}
\Psi(\X,\Y)=d\mphip\;\phineu(\X,\Y)-\frac{d}{2}\sphip\;\Phis(\X,\Y)+C,
\eeq 
where $\phineu$ and $\Phis$ are elementary problems, sketched in figure \ref{Fig4}, satisfying the Laplace equation \eqref{Laplace} and with the following asymptotic  behaviour
\beq\label{ag1}
\toutind
\phineu\underset{\X\to\pm\infty}{\sim}\X\pm \cA/2,\quad & \phineu\underset{\Y\to+\infty}{\sim} 0,\\
\Phis\underset{\X\to\pm\infty}{\sim} \mp \X,\quad & \Phis\underset{\Y\to+\infty}{\sim} \frac{\Y}{\th}+\cAs,
\toutout
\eeq
($\phineu$ and $\Phis$ are defined up to  constants and we have chosen an antisymmetric form for  $\phineu$ and a symmetric form for $\Phis$). In the above, the parameters  $(\cA,\cAs)$  which depend on $\th$ only  are to be determined. 

We now identify \eqref{psi} for $\X\to\pm\infty$ with \eqref{eq:3.5}   for $|x-\xgp| \to 0$,  and  \eqref{psi}  for $\Y\to +\infty$ with \eqref{eq:3.10} for $y\to 0$. In doing so, we obtain 
\beq\nonumber
\phig(\xgp^\pm)=\pm \frac{\cA}{2}\,d\mphip+C,\quad \phi_1(0)=-\frac{\cAs}{2}\,d\sphip+C,
\eeq
and eliminating $C$ from the above expressions, 
\beq\label{rr00}
 \toutin
\dsp\sphi= \cA \,d\mphip,
\vspace{.1cm}\\
\dsp \phi_1(0)-\mphi=
\cAs\, a \phi_1'(0),
\vspace{.1cm}\\
\dsp d\sphip
 +2a \phi_1'(0)=0.
 \toutout
 \eeq
 
 \vspace{.3cm}
  The jump conditions  \eqref{rr00} are the main results of our analysis. They apply whatever    channel lengths and end conditions are applied (provided the underlying assumptions that $kd,ka \ll 1$ are maintained).
   Besides,
 the parameters $(\cA,\cAs)$ are explicit (see appendix \ref{app}), specifically
\beq
\label{defAAs}
\toutin
\cA=\frac{2}{\pi}\left(
\log (1+\th^2)-2\th \tan^{-1}\th
\right),\vspace{.2cm}\\
\cAs=\frac{2}{\pi}\left(
\log \frac{(1+\th^2)}{4\th}+(1/\th-\th) \tan^{-1}\th
\right).
\toutout
\eeq

\subsubsection{Solution in the junction region with two  channels}
When the distance between two channels is subwavelength,  the junction contains both
side channel openings. We repeat the same exercice as previously with now, the conservation of the flux being $\sphip+2\th(\phi_1'(0)+\phi'_2(0))=0$. Also, for convenience, we define
\beq\label{as}
\phics=\phi_1+\phi_2,\qquad \phica=\phi_1-\phi_2,
\eeq
being the symmetric and antisymmetric fields in the channels. The matching conditions on  $\grad\Psi$ read
\beq\label{bc2}\toutin
\grad\Psi\underset{\X\to\pm\infty}{\sim} d\phig'(\xgp^\pm)\ex=d\left(\mphip\pm\frac{1}{2}\sphip\right)\ex,\vspace{.2cm}\\
\grad\Psi\underset{\Y\to+\infty_1}{\sim} d\phi_1'(0)\ey=-\frac{\ey}{4\th}\,d\sphip+\frac{\ey}{2}\,d\phica'(0),\vspace{.2cm}\\
\grad\Psi\underset{\Y\to+\infty_2}{\sim} d\phi_2'(0)\ey=-\frac{\ey}{4\th}\,d\sphip-\frac{\ey}{2}\,d\phica'(0).
\toutout\eeq
The system \eqref{Laplace} with \eqref{bc2} is linear with respect to $\mphip$, $\sphip$ and $\phica'(0)$ from  which we deduce that $\Psi$ can be sought of the form 
\beq\label{psi2}
\Psi=d\mphip\;\phineu-\frac{d\sphip}{2}\;\Phis+a\phica'(0)\,\Phia+C,
\eeq 
with $\phineu$, $\Phis$ and $\Phia$ satisfying the Laplace equation and with the following asymptotic behaviours 
\beq\nonumber\toutin
\phineu\underset{\X\to\pm\infty}{\sim}\X\pm \cB/2,\quad  \phineu\underset{\Y\to\infty_1}{\sim} - \cBac/2, \quad \phineu\underset{\Y\to\infty_2}{\sim}  \cBac/2,\vspace{.3cm}\\
\begin{array}{ll}
\Phis\underset{\X\to\pm\infty}{\sim} \mp \X, & \Phis\underset{\Y\to+\infty_1}{\sim} \frac{\Y}{2\th}+\cBs/2,\\
&   \Phis\underset{\Y\to+\infty_2}{\sim} \frac{\Y}{2\th}+\cBs/2,\vspace{.3cm}\\
\Phia\underset{\X\to\pm\infty}{\sim} \mp \cBac/2, & \Phia\underset{\Y\to+\infty_1}{\sim} \pm\frac{\Y}{2\th}\pm\cBa/2,\\
& \Phia\underset{\Y\to+\infty_2}{\sim} \pm\frac{\Y}{2\th}\pm\cBa/2,
\end{array}
\toutout
\eeq
($\Y\to+\infty_1$ means $\Y\to+\infty$ in the first channel, $\Y\to+\infty_2$ means $\Y\to+\infty$ in the second channel).
  \begin{figure}[h!]
\centering
\includegraphics[width=1\columnwidth]{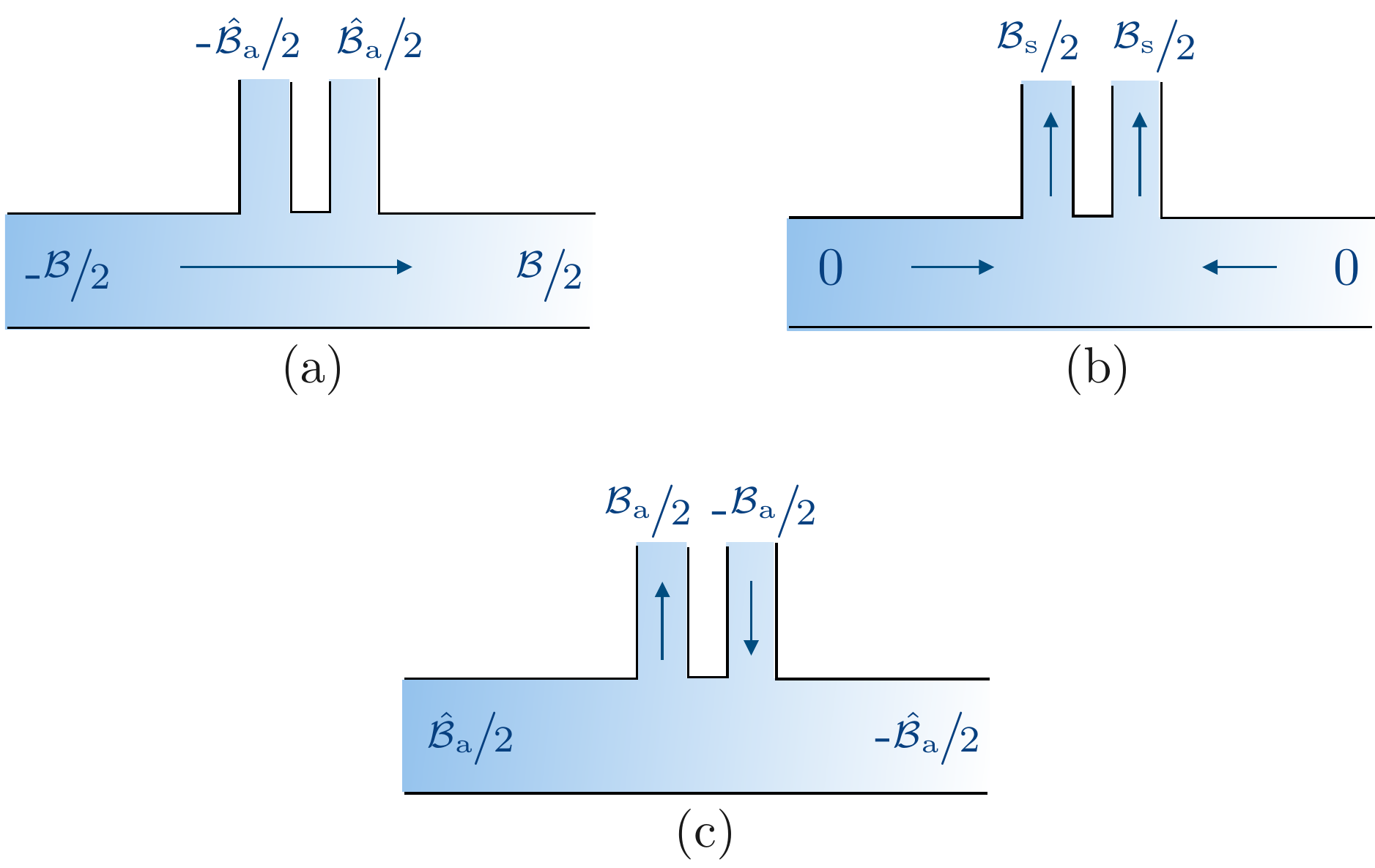}
	\caption{ Elementary problems of potential flows with \eqref{bc2} set on (a) $\phineu$,
	 (b) $\Phis$ and (c) $\Phia$.}
	\label{Fig5}
\end{figure}
Note that the parameters $\cB$ and $\cBs$ are associated with similar elementary problems as those appearing for a single channel, see (a) and (b) in figures \ref{Fig4} and \ref{Fig5}. In contrast  the parameters  $\cBa$ and $\cBac$ are associated with a new antisymmetric problem  involving the exchange of flux between the two channels  (which was obviously  not possible for a single channel).
The parameter $\cBac$ appears in the two problems on $\phineu$ and $\Phia$; indeed, integrating by part the product  $\phineu\Delta\Phia=0$,  we obtain $\cBac=-\int \grad\phineu\grad\Phia$ using 
that $\phineu{\sim} \mp \cBac/2$ for $\Y\to \infty$ in the channels; in doing the same for
$\Phia\Delta\phineu=0$, we obtain $\cBac=-\int \grad\phineu\grad\Phia$ using that $\Phia{\sim} \mp \cBac/2$ for $\X\to\pm\infty$ in the guide.  

\vspace{.2cm}
As for a single channel, we  now identify  \eqref{psi2}  for $\X\to\pm\infty$ with \eqref{eq:3.5}   for $|x-\xgp| \to 0$,  and  \eqref{psi2}  for $\Y\to +\infty_n$, $n=1,2$ with \eqref{eq:3.10} for $y\to 0$ in the first channel and in the second channel. 
This results in 
\beq\nonumber\begin{array}{l}
\phig(0^\pm)=\pm\frac{\cB}{2}\,d\mphip\mp\frac{\cBac}{2}a\phica'(0)+C,\vspace{.1cm}\\
\phi_1(0)=-\frac{\cBac}{2}d\mphip-\frac{\cBs}{2}\frac{d\sphip}{2}+\frac{\cBa}{2}a\phica'(0)+C,\vspace{.1cm}\\
\phi_2(0)=\frac{\cBac}{2}d\mphip-\frac{\cBs}{2}\frac{d\sphip}{2}-\frac{\cBa}{2}a\phica'(0)+C,
\end{array}\eeq 
and thus
\beq\label{rr0}
 \toutin
\dsp\sphi= \cB\,d\mphip -\cBac \,a\phica'(0),
\vspace{.1cm}\\
\dsp \phics=2\mphi+\cBs\,a\phics',
\quad
\dsp \phica=-\cBac\, d\mphip+\cBa \,a\phica',
\vspace{.1cm}\\
\dsp d\sphip
 +2a\phics'=0.
 \toutout
 \eeq

  The jump conditions  \eqref{rr0} are the   jump conditions produced by two channels with subwavelength distance (as opposed to  \eqref{rr00}).
As for a single channel, they apply whatever the heights of the two channels and whatever the end conditions are (and the heights and end conditions of the two channels can be different). 

\subsection{The final models }
The final models are now written  considering that the channels have closed ends at $y=b$, hence 
\beq \label{phii} \phi_n(y)=A_n\frac{\cos k(y-b)}{\cos k b},\quad n=1,2.\eeq
In the model a for a single channel,  we thus  have
$\phi_1(0)=A_1$ and $\phi_1'(0)=A_1 k\tan kb$
which can be used in  \eqref{rr00} to determine the jump conditions applying on $\phig$ only,
after eliminating  $A_1$. They read
\beq\label{saut1}
\toutin
 \dsp d [\phig']=\als\overline{\phig}, \quad  [\phig]=\ala\, d\overline{\phig'},\vspace{.3cm}\\
 \als=\frac{2\X}{\cAs\X-1},\qquad 
\ala=\cA,
\toutout\eeq
where   $\X=ka\tan kb$.
 Once $\phig(x)$ has been obtained,  $A_1$ can be determined from \eqref{rr00}.
In the model b, 
\eqref{phii} applies to $\phis$ and $\phia$  with  $A_\text{\tiny s}=(A_1+A_2)$, $A_\text{\tiny a}=(A_1-A_2)$, from  \eqref{as}. Eliminating $A_\text{\tiny s}$ and $A_\text{\tiny a}$ in \eqref{rr0} yields 
\beq\label{saut2}\toutin
 d [\phig']=\bes\overline{\phig}, \quad  [\phig]=\bea\, d\overline{\phig'}.
\vspace{.3cm}\\
 \bes=\frac{4\X}{\cBs\X-1},\qquad
\bea=\cB-\frac{\cBac^2\X}{\cBa\X-1}.
\toutout\eeq
Eventually, once $\phig(x)$ has been obtained, $(A_\text{\tiny s},A_\text{\tiny a})$ (and $(A_1,A_2)$) can be obtained from \eqref{rr0} (and \eqref{as}).

 \section{Scattering by a single channel}
 \label{S3}
 
 \begin{figure}[b!]
\centering
\includegraphics[width=1\columnwidth]{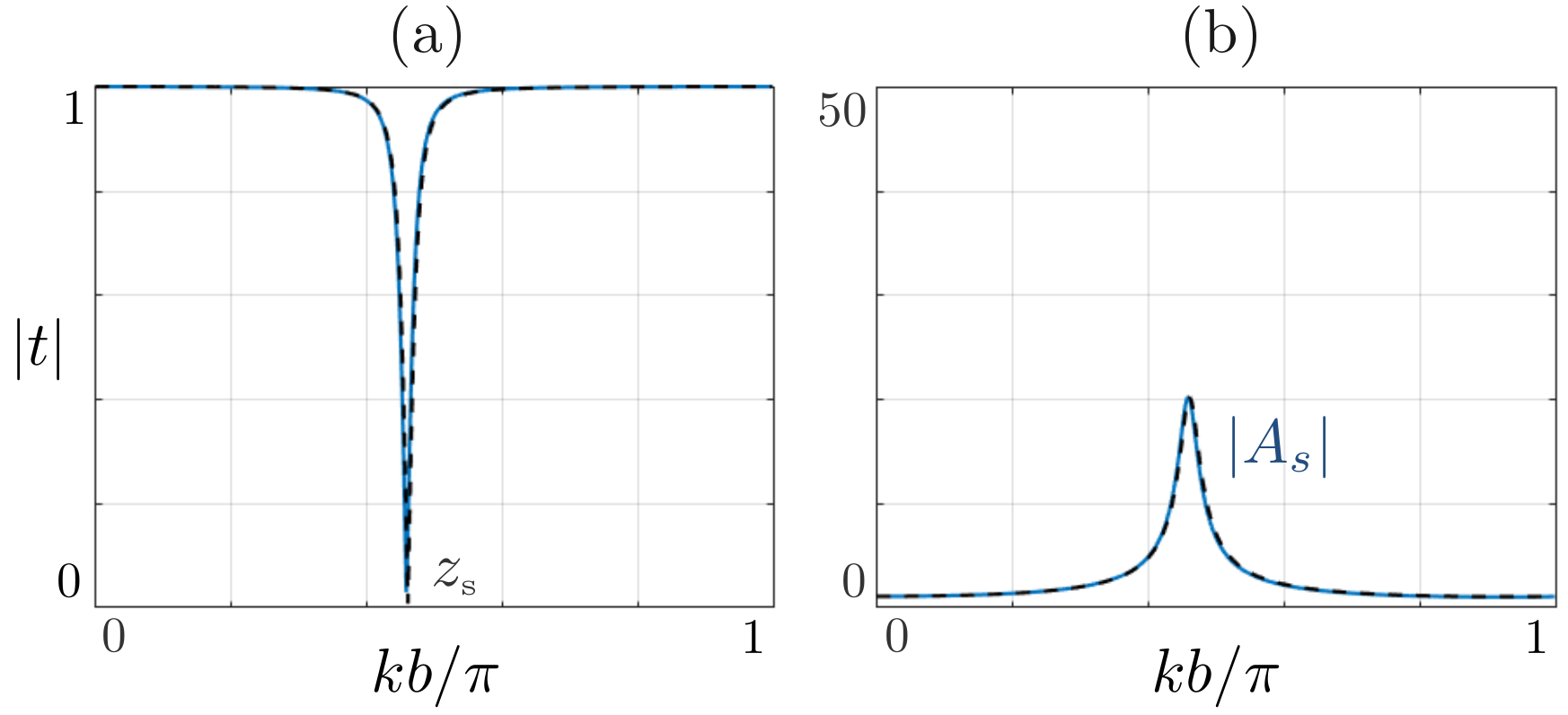}
	\caption{
	(a) Transmission $|t|$ by a single channel and (b) amplitude $|\As|$ at the top of the channel (here $\As=A_1$) against the non-dimensional frequency $kb/\pi$ ($d=b=20a$). Direct numerics (plain lines) and from \eqref{rt} (dashed black lines).}
	\label{Fig6}
\end{figure}
To begin with, we consider a single channel opening onto the  main guide at $x=0$.
 The features of this problem will provide an important guide to the discussion of two channels.
  For an incident right-going wave, we look for a solution $\phig(x)$ of the form
\beq\toutind
\dsp \phig(x)=e^{ikx}+r e^{-ikx},& x\in(-\infty,0),\\
\dsp  \phig(x)=t e^{ikx},\quad & x\in(0,+\infty),
\toutout\nonumber
\eeq
with the jump conditions \eqref{saut1}  at $\xgp=0$. We obtain
\beq\label{rt}\toutind
\dsp r=\frac{1}{2}\left(\frac{\ys}{\ys^*}-\frac{\ya}{\ya^*}\right),\quad &
\dsp t=\frac{1}{2}\left(\frac{\ys}{\ys^*}+\frac{\ya}{\ya^*}\right),\vspace{.2cm}\\
\dsp \ys=1-\frac{i\als}{2kd},\quad &
\dsp \ya=1+\frac{ikd \ala}{2},
\toutout\eeq
with $(\als,\ala)$ defined in \eqref{saut1} along with \eqref{defAAs}.
As it should be, the scattering coefficients satisfies the energy conservation $|r|^2+|t|^2=1$ and the reciprocity $r t^*+tr^*=0$. The figure \ref{Fig6} shows the transmission $|t|$ and the amplitude $A_1=\als/(\X\ys^*)$  at the top of the channel for $d=b=20a$.

\vspace{.2cm}
The effect of this thin channel is in general weak (hence $|t|\simeq1$) except at  the quarter-wavelength resonance for $kb\simeq \pi/2$ which produces a zero-transmission ($z_\text{\tiny s}$).  
From \eqref{rt} this first zero-transmission corresponds to $\text{cot}k\beff=0$ 
with  $\beff$   the so-called effective length  given by
\beq\label{bad}\begin{array}{l}
\ksz=\frac{\pi}{2\beff},\qquad \beff=(b+\bad),\vspace{.2cm}\\
\bad=\frac{2}{\pi}\left(\frac{1}{\th}\tan^{-1}\th-\log\left(\frac{4\th}{\sqrt{1+\th^2}}\right)\right)\;a.
\end{array}\eeq

\begin{figure}[h!]
\centering
\includegraphics[width=1\columnwidth]{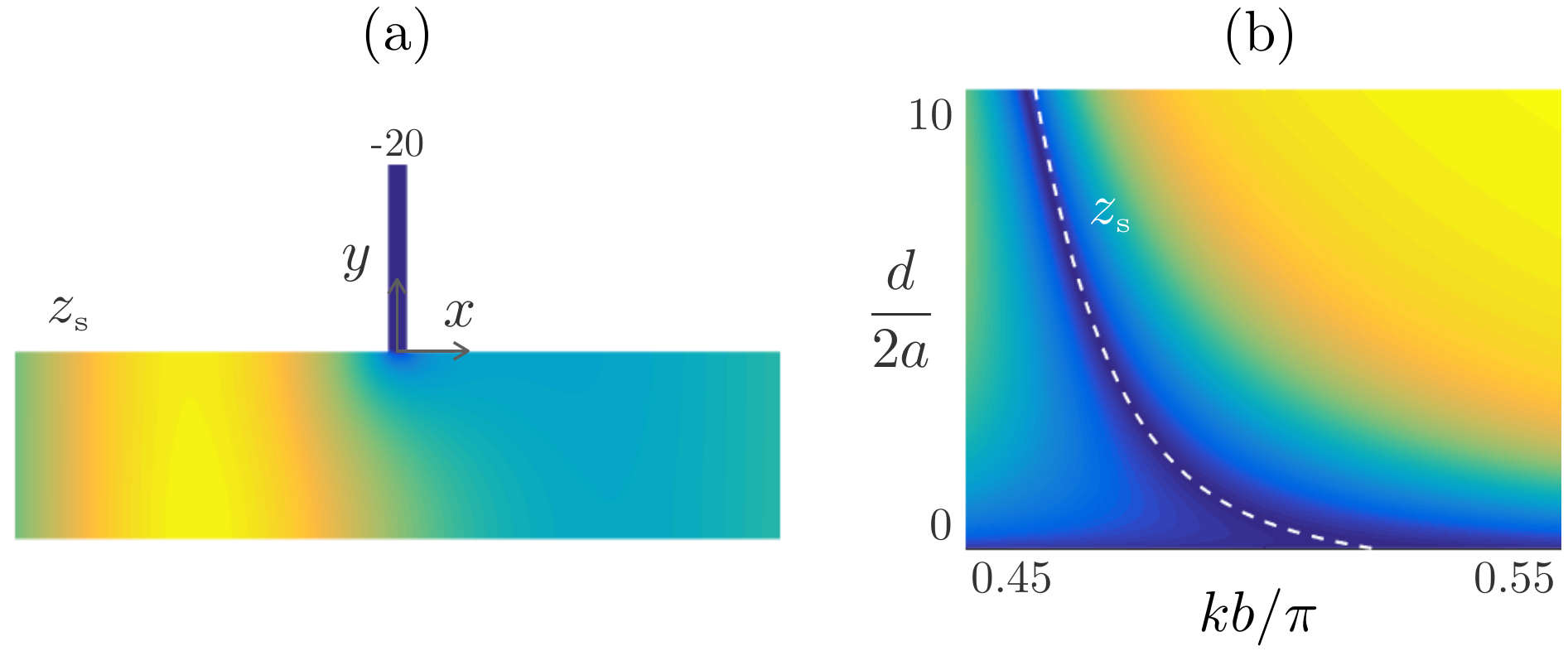}
	\caption{ (a) Real part of the field at the resonance of a single channel, (b) Transmission coefficient $|t|$  from direct numerics against non-dimensional frequency $kb/\pi$ and non-dimensional guide width $d/2a$. The dashed white line shows $\ksz$ from \eqref{bad}. }
	\label{Fig7}
\end{figure}

The  variation of $\ksz$ against $d/a$ is  manifested in   the zero of $|T|$ in the $(kb/\pi,d/2a)$ plane in figure \ref{Fig7}(b) (the dashed white line shows $k_0$ in \eqref{bad}; the relative error w.r.t. the numerics is less than 0.05\% up to $d/2a=5$ afterwards it increases slightly, up to 0.3\% at $d/2a$=10). 
 We did not find a similar theoretical prediction for the added length in the literature. However, we notice the good agreement between \eqref{bad} and the fits proposed in \cite{cervenka2018,cervenka2019} resulting in  $\frac{\bad}{2a}=0.52$  for $\theta=\frac{2}{15}$ (\eqref{bad}  provides $\frac{\bad}{2a}=0.5193$).

Finally, and anticipating an easier comparison with the case of two channels, we provide an approximate expression of $t$  in \eqref{rt}  close to the resonance for $k\beff=\frac{\pi}{2}+\varepsilon$, $\varepsilon\ll 1$ (hence $\X\simeq -\frac{\pi a}{2\beff \varepsilon}$ in \eqref{saut1}). In doing so, we obtain
\beq\label{polert}
t=\frac{(k-\ksz)}{(k-\ks)},\quad \ks=\ksz-i\frac{\th}{b},
\eeq
 with $\ksz$, in  \eqref{bad}, the wavenumber at zero-transmission.

 \section{Close channels or far channels, two different resonant mechanisms}
  \label{S4}
 
\subsection{Far channels, AIT in the classical model a}
In the classical model a,  the evanescent fields are confined to the 
vicinity of each  channel with no overlap. Hence, the solution is sought in the form
\beq\label{solfar}\toutind
\dsp \phig(x)=e^{ikx}+R e^{-ikx},\quad & x\in(-\infty,-h/2),\vspace{.1cm}\\
\dsp \phig(x)=\Gs\cos kx+\Ga \sin kx,& x\in(-h/2,h/2),\vspace{.1cm}\\
\dsp  \phig(x)=T e^{ikx},\quad & x\in(h/2,+\infty),
\toutout
\eeq
with \eqref{saut1} applying  at $\xgp=\pm h/2$. It follows that
\beq\label{RTa}
\toutin
\dsp R=\frac{1}{2}\left(\frac{\Zs}{\Zs^*}-\frac{\Za}{\Za^*}\right),\quad 
\dsp T=\frac{1}{2}\left(\frac{\Zs}{\Zs^*}+\frac{\Za}{\Za^*}\right),\vspace{.4cm}\\
\dsp \Zs=(1+i\ta)\left(1+\frac{\ala\als}{4}\right)-i\frac{\als}{kd}-\ala kd\, \ta,\vspace{.3cm}\\
\dsp \Za=(1+i\ta)\left(1+\frac{\ala\als}{4}\right)+\frac{\als}{kd}\ta+i\ala kd, 
\toutout\eeq
where $\ta=\tan (kh/2)$.
\vspace{.2cm}

\begin{figure}[t!]
\centering
\includegraphics[width=1\columnwidth]{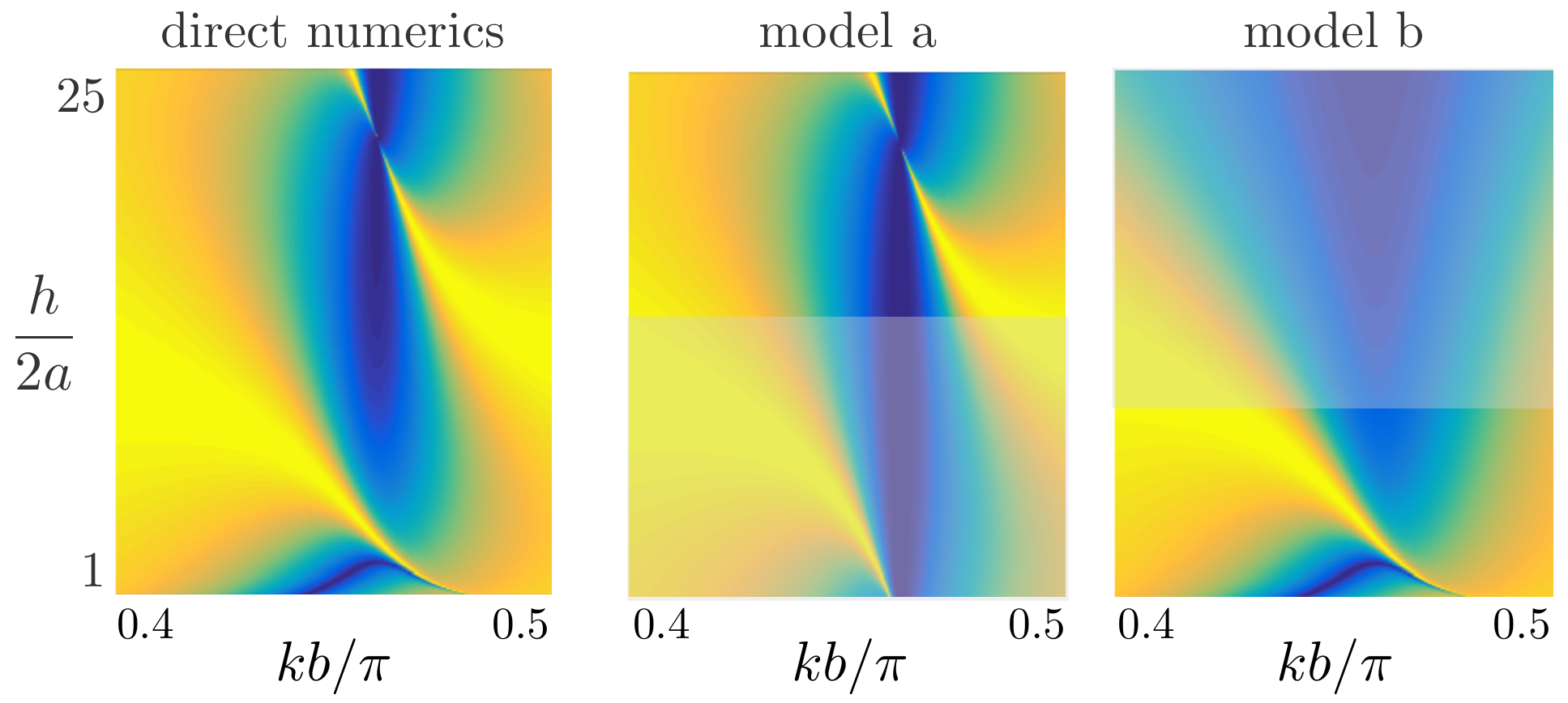}
	\caption{ Comparison of the transmission spectra given by direct numerics and  given by the model a, from \eqref{RTa} and by the model b, from \eqref{RTb}. The grey regions  hide the region where each model is not valid ($d=b=20a$).}
	\label{Fig8}
\end{figure}

 The  transmission spectrum given by \eqref{RTa} in the model a is shown in figure \ref{Fig8} to be compared 
  to that calculated numerically  (the same as in figure \ref{Fig2}(a)).
As   expected, when the two channels have  large separation distance $h$, the interaction between them (when they interact) is attributable to interference of propagating waves, hence the model a, which is the classical model, is valid. 
  \begin{figure}[b!]
\centering
\includegraphics[width=1\columnwidth]{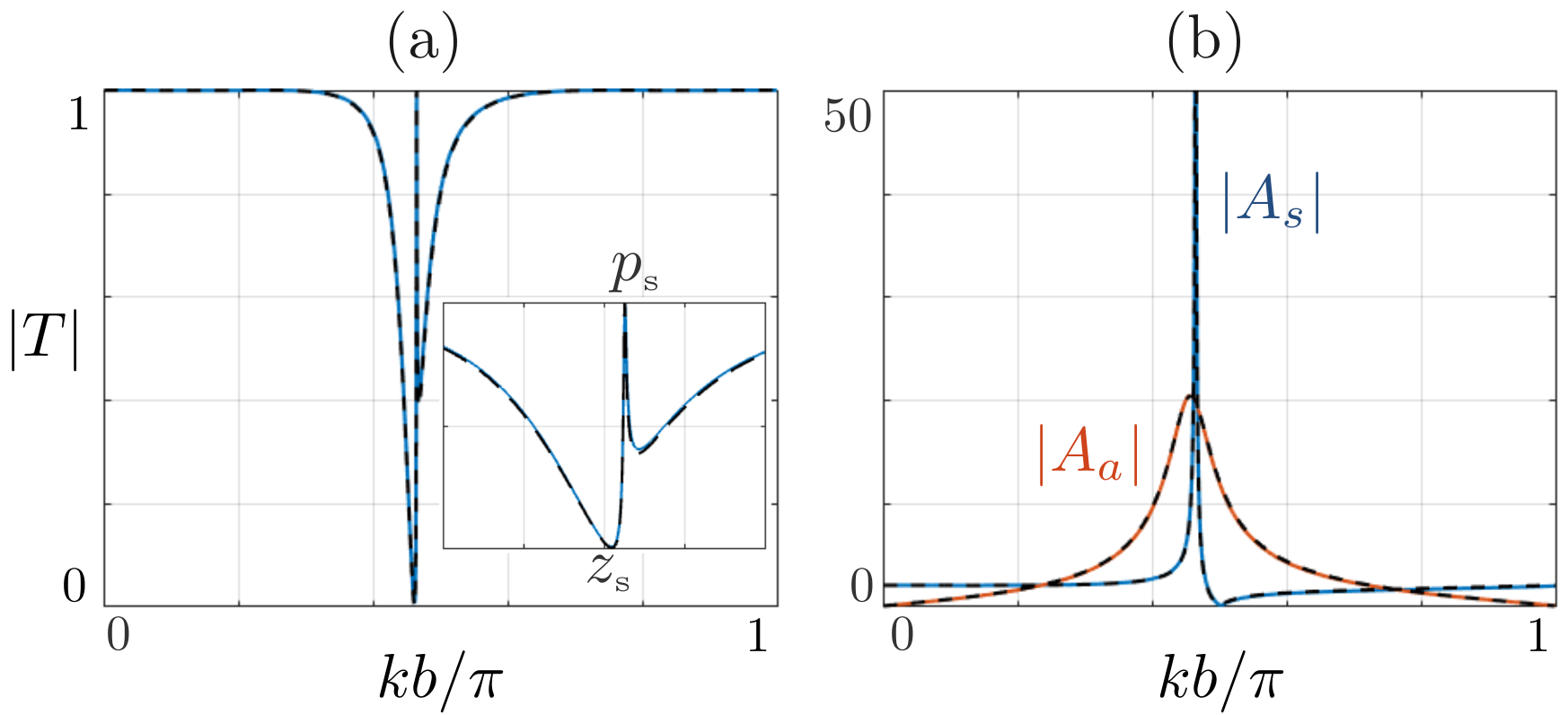}
	\caption{(a) Transmission  against   $kb/\pi$; the inset shows the range $kb/2\pi\in (0.42,0.5)$ (b) Symmetric $\As$ and antisymmetric $\Aa$ amplitudes  ($h=2b$ and  $b=d=20a$). Direct numerics (plain lines) and from \eqref{RTa} (dashed black lines).}
	\label{Fig9}
\end{figure}
We provide  quantitative comparisons  in figure \ref{Fig9}  where we report  the variations of   $|T|$ and of $(\As,\Aa)$ for $h=20a$. 
The transmission resembles that  obtained for a single channel except for the occurrence 
 of a perfect transmission (named $\ps$) following the zero-transmission $\zs$, a feature characteristic of 
  Fano resonance and which is accurately captured by the model.  The shape of the quasi-trapped associated with the Fano resonance  is visible at $\ps$  in fi\-gu\-re \ref{Fig10}.

 \begin{figure}[h!]
\centering
\includegraphics[width=.9\columnwidth]{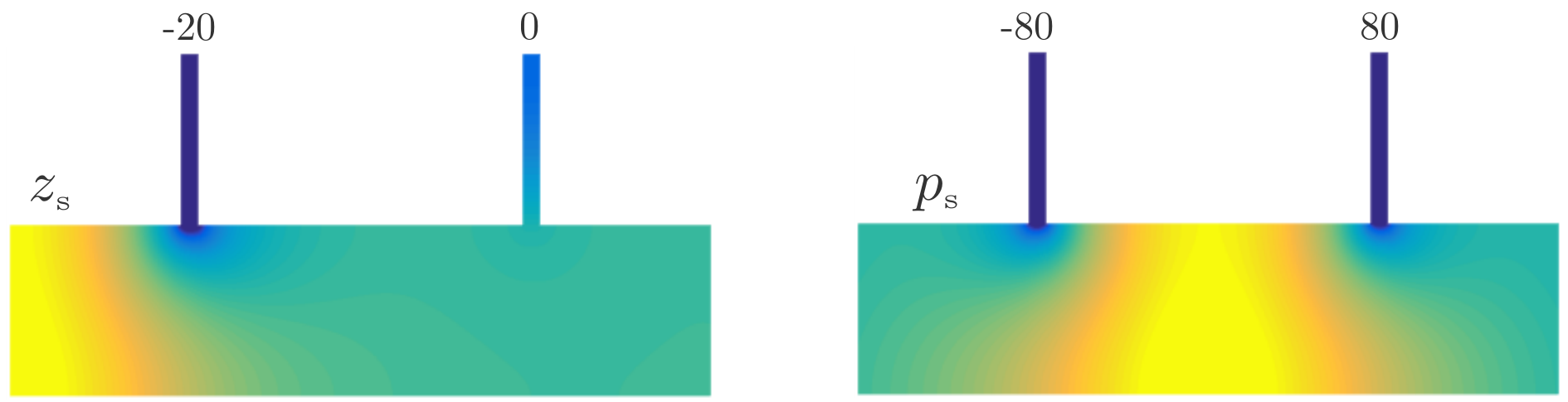}
	\caption{Real parts of the fields at the zero-transmission ($\zs$) and at the perfect transmission ($\ps$)  revealing the shape of the symmetric quasi-trapped mode.}
	\label{Fig10}
\end{figure}

 The complex poles of the Fano resonance and the associated perfect- or quasi-trapped modes have been analyzed in several studies see, {\em e.g.}    \cite{chesnel2018}. Our effective model  provides  approximate expressions of these poles  owing to  expansions of the scattering coefficients  in the vicinity of $k\beff=\pi/2$ as we have done for a single channel. We 
define  the detuning parameter $\delta$ as
\beq\nonumber
\delta=\pi\left(\frac{h}{2\beff}-1\right),
\eeq
hence for  $\delta=0$, the Bragg and the quarter-wavelength resonances are perfectly tuned. 
From \eqref{RTa}, we obtain 
\beq\label{Tappa}
T\simeq \frac{(k-\ksz)^2}{(k-k^+)(k-k^-)},
\eeq
with $\ksz$  defined in \eqref{bad} and 
\beq\begin{array}{l}
\dsp k^-=k_0+\frac{\th\delta}{\beff}-i\frac{2\th}{\beff},\quad 
\dsp k^+=k_0-\frac{\th\delta}{\beff}-i\frac{\th\delta^2}{2\beff}.
\end{array}
\eeq
The two poles are dictated by the detuning parameter with the pole $k^-$ relating to a weak resonance    correspon\-ding roughly to the pole of a single channel of width $4a$ and the pole $k^+$ relating to the strong resonance  associa\-ted with the quasi-trapped mode resonance. In particular, when $\delta=0$, $k^+=k_0$ becomes real, the quasi-trapped mode becomes a perfect-trapped mode and the Fano re\-so\-nan\-ce disappears.

 \subsection{Close channels, the model b}
 Contrary to the model a,  model b concerns the case where the channels are close enough  that the evanescent field  in the junction region  plays a leading order effect in connecting the two channels. This is accounted for in the model b by reducing the region of the two channels to a single junction. Accordingly, we look for a solution $\phig(x)$
\beq\toutind
\dsp \phig(x)=e^{ikx}+R e^{-ikx},& x\in(-\infty,0),\\
\dsp  \phig(x)=T e^{ikx},\quad & x\in(0,+\infty),
\toutout
\eeq
and we apply the jump conditions \eqref{saut2} at the single junction $\xgp=0$. The resulting scattering coefficients hence read 
\beq\label{RTb}\toutind
\dsp R=\frac{1}{2}\left(\frac{\zss}{\zss^*}-\frac{\zaa}{\zaa^*}\right),\quad &
\dsp T=\frac{1}{2}\left(\frac{\zss}{\zss^*}+\frac{\zaa}{\zaa^*}\right),\vspace{.2cm}\\
\dsp \zss=1-\frac{i\bes}{2kd},\quad &
\dsp \zaa=1+\frac{ikd \bea}{2}, 
\toutout\eeq
with $(\bes,\bea)$ in \eqref{saut2}. It is worth noticing that  the effect of $h$ is now entirely encapsulated in the effective parameters $(\cBs,\cBa,\cBac)$. Furthermore, neglecting the evanescent field (resulting in $\bes=-4\X$ and $\bea=0$ in \eqref{saut2}) would completely fail  in predicting the observed spectrum. Indeed, doing so would reduce the two channels to a single one.

\vspace{.3cm}

\begin{figure}[h!]
\centering
\includegraphics[width=1\columnwidth]{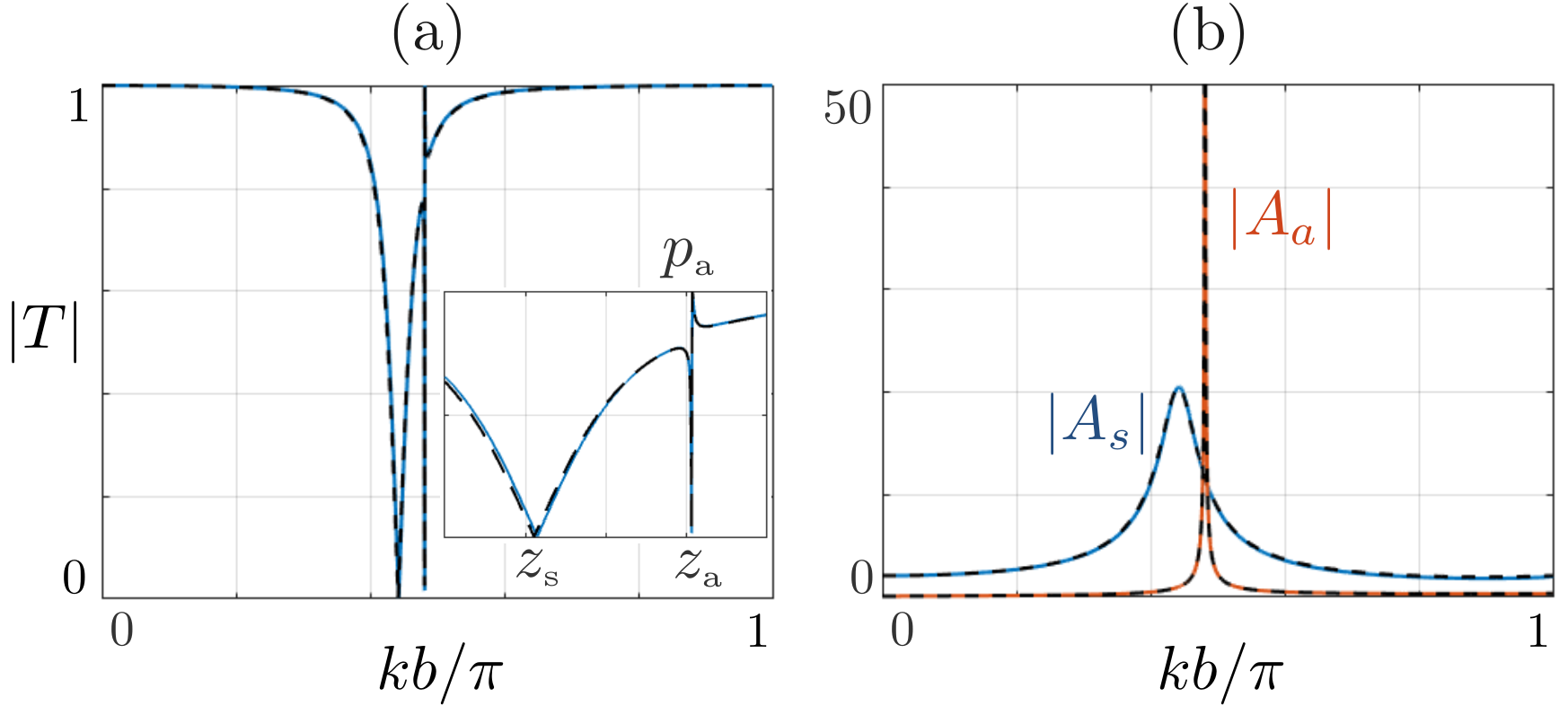}
	\caption{Transmission $|T|$ and amplitudes $(\As,\Aa)$, same representation as in figure \ref{Fig9} for  $h=a$ ($b=d=20a$); the maximum of $\Aa$ is at about 300.}
	\label{Fig11}
\end{figure}

\begin{figure}[h!]
\centering
\includegraphics[width=1\columnwidth]{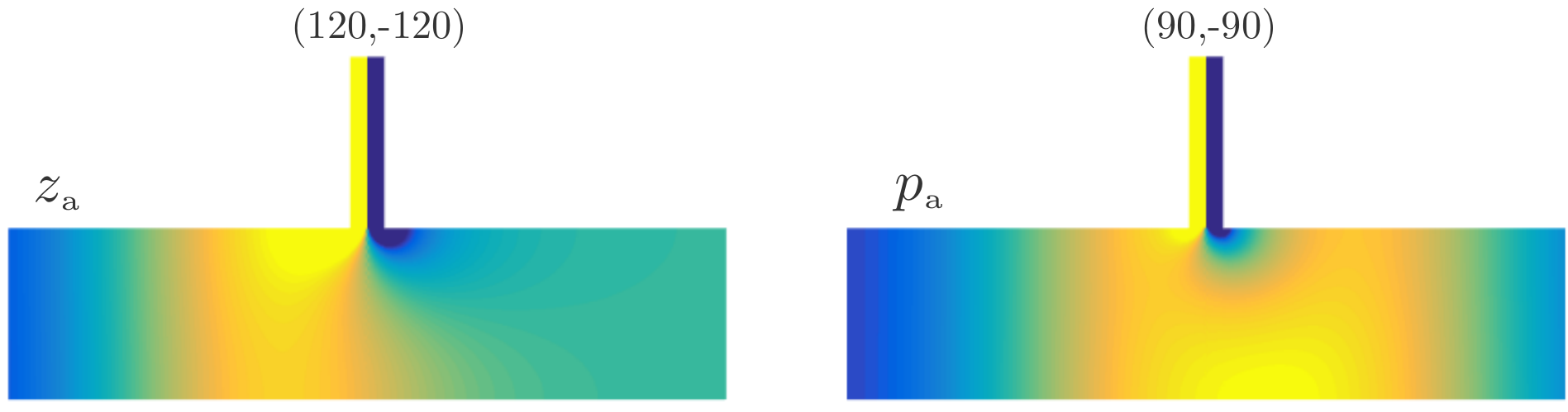}
	\caption{Real part of the fields at the zero-transmission $\za$ and the perfect-transmission $\pa$  associated with the dipolar behavior of the pair of contiguous channels. }
	\label{Fig12}
\end{figure}

To begin with, we come back to the figure \ref{Fig8}  
where, in panel (c), we showed 
 the transmission spectrum given by \eqref{RTb}. As expected,  model b is valid only for small distances, $h$,  which  include the region of the Autler-Townes splitting. Eventually, the model
 a (resp. b)  is valid  for $h/2a\gtrsim 7$ (resp.  $h/2a \lesssim 7$). This corresponds to $kh\sim 1$ which is consistent with the asymptotic analysis presented in \S \ref{S2}.
The accuracy of the model b is  further illustrated in fi\-gu\-re \ref{Fig11} where we report  $|T|$ and $(\As,\Aa)$ against the non-dimensional frequency for $h=a$ (the two channels are contiguous). The resonance due to ATS is associated with a strong dipolar behaviour of the two channels, able to produce a zero-transmission (named $\za$) immediately followed by a perfect-transmission (named $\pa$). The corresponding patterns are  shown in figure \ref{Fig12} where we see the strong evanescent field inducing a dipolar behaviour in the vicinity of the channel openings.


\vspace{.2cm}
Contrary to the case of the Fano resonance, the re\-so\-nance associated with ATS has received much less attention. In \cite{cheng2019}, the dipolar behavior of the ATS resonance has been observed for two Helmholtz resonators  (see figure 4 (c-d) in this reference), and  the Akaike criterion has been used to discriminate the ATS from the AIT. Close to the resonance at $kb=\pi/2$, and using Taylor expansions as before, we obtain
\beq\label{Tappb}
T\simeq \frac{(k-k_0^+)(k-k_0^-)}{(k-k^+)(k-k^-)},\quad
\eeq
with $k_0^\pm$ associated with zero-transmissions   given by
\beq\label{K0}
k_0^\pm-\ksz=\pm\frac{\pi a}{2b^2}\sqrt{\Delta},
\eeq
(and $\ksz$ in \eqref{polert})
and with the complex poles $k^\pm$  given by
\beq\label{gg1}\toutin
 k^+-\ksz=\frac{\pi a}{2b^2}\delta_r-2i\frac{\th}{b}\left(\frac{\pi  d}{4b} \cBac\right)^2
\vspace{.2cm}\\
 k^--\ksz=-\frac{\pi a}{2b^2}\delta_r-2 i\frac{\th}{b},
\toutout\eeq
(with $\ks$  in \eqref{polert}). In the above, 
we have defined $\Delta=(\delta_r^2-\cBac^2)$ and  $2\delta_r=(\cBs-\cBa)$
and we have  used that  $\cB\simeq 2\cA$ and that $(\cBs+\cBa)\simeq 2\cAs$; see appendix \ref{app2}. 
\begin{figure}[b!]
\centering
\includegraphics[width=1\columnwidth]{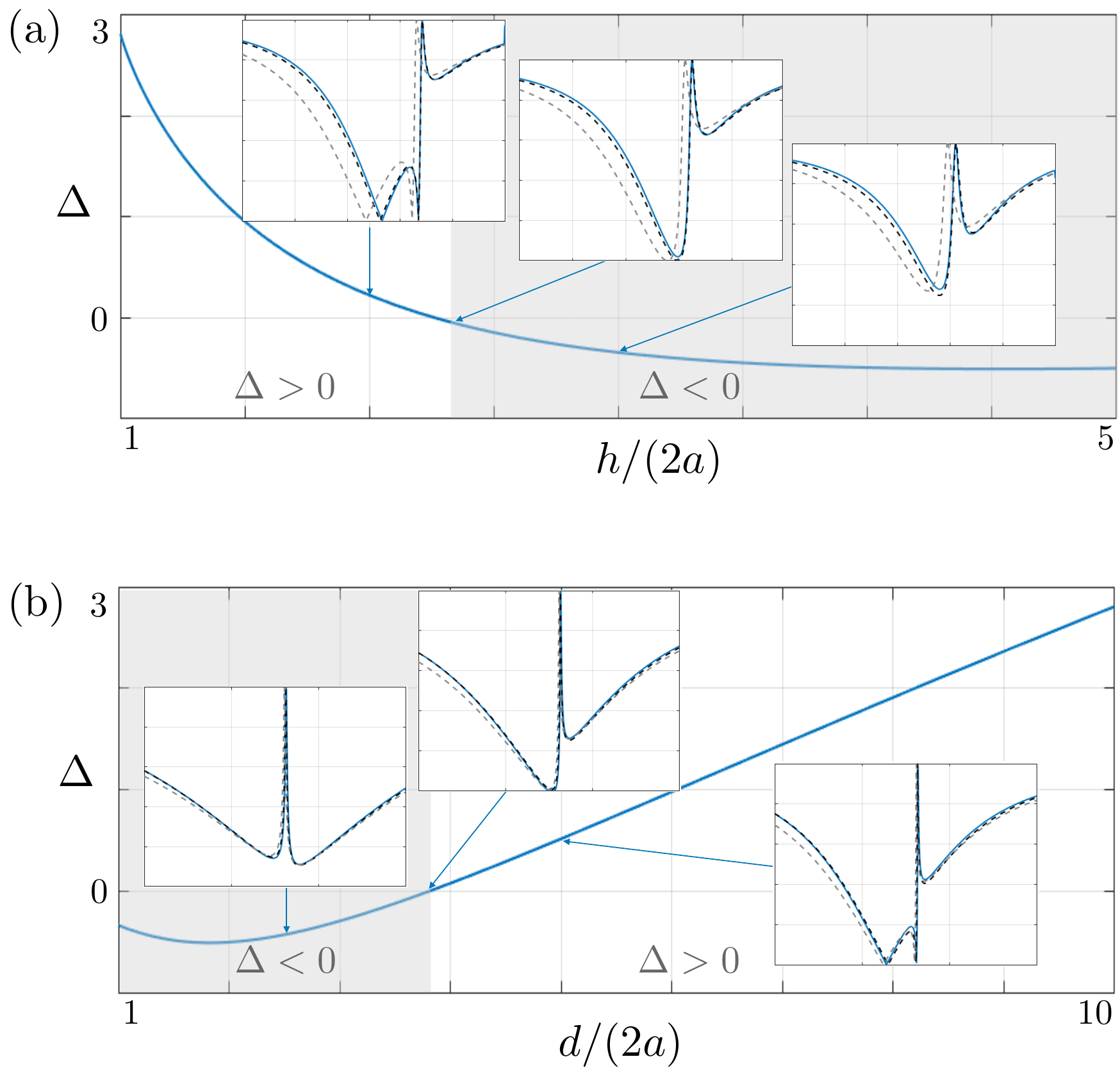}
	\caption{Variations of $\Delta$ in \eqref{K0} against (a) the non dimensional spacing $h/2a$   ($d=10a$), and (b) the non dimension guide width $d/2a$  ($h=a$) (b). The insets show $|T|$ against $kb/\pi\in(0.4,0.5)$ in (a) and  $kb/\pi\in(0.4,0.55)$ in (b) from direct numerics (plain lines) from $T$ in \eqref{RTb} (black dashed lines) and its approximate version \eqref{Tappb} (grey dashed lines). }
	\label{Fig13}
\end{figure}
The different shapes  in the transmission profiles are shown in figure \ref{Fig13} for $\Delta>0$ resulting in two zero-transmissions, $\Delta=0$ resulting in a single zero-transmission and $\Delta<0$ resulting in no zero-transmission. 
In all cases, the model b  captures these variations accurately using $T$ in \eqref{RTb} (dashed black lines)  and reasonably using the approximate form in \eqref{Tappb} (dashed grey lines).

\vspace{.3cm}
 We have seen that the  single zero-transmission of the Fano resonance  occurs for $k=\ksz$, associated with that of a single channel (from \eqref{Tappa}). Here, in contrast, the two zero-transmissions  are split symmetrically with respect to   $\ksz$ along the real axis when $\Delta>0$ and along the imaginary axis when $\Delta<0$; the same applies for the real parts of two poles.  However, the relative positions of the complex parts  of the poles depend on the geometry notably through $\cBac$. With $k^-$ associated with the symmetric solution and $k^+$ with the antisymmetric solution, the pole $k^+$ is, in general, closer to the real axis. The most striking feature which allows us to eliminate   the Fano resonance is that there is no trapped mode when the channel are sufficiently close and this is consistent with the fact that $\cBac^2>0$ (see appendix \ref{app2}).

 \vspace{.3cm}

\begin{figure}[h!]
\centering
\includegraphics[width=1\columnwidth]{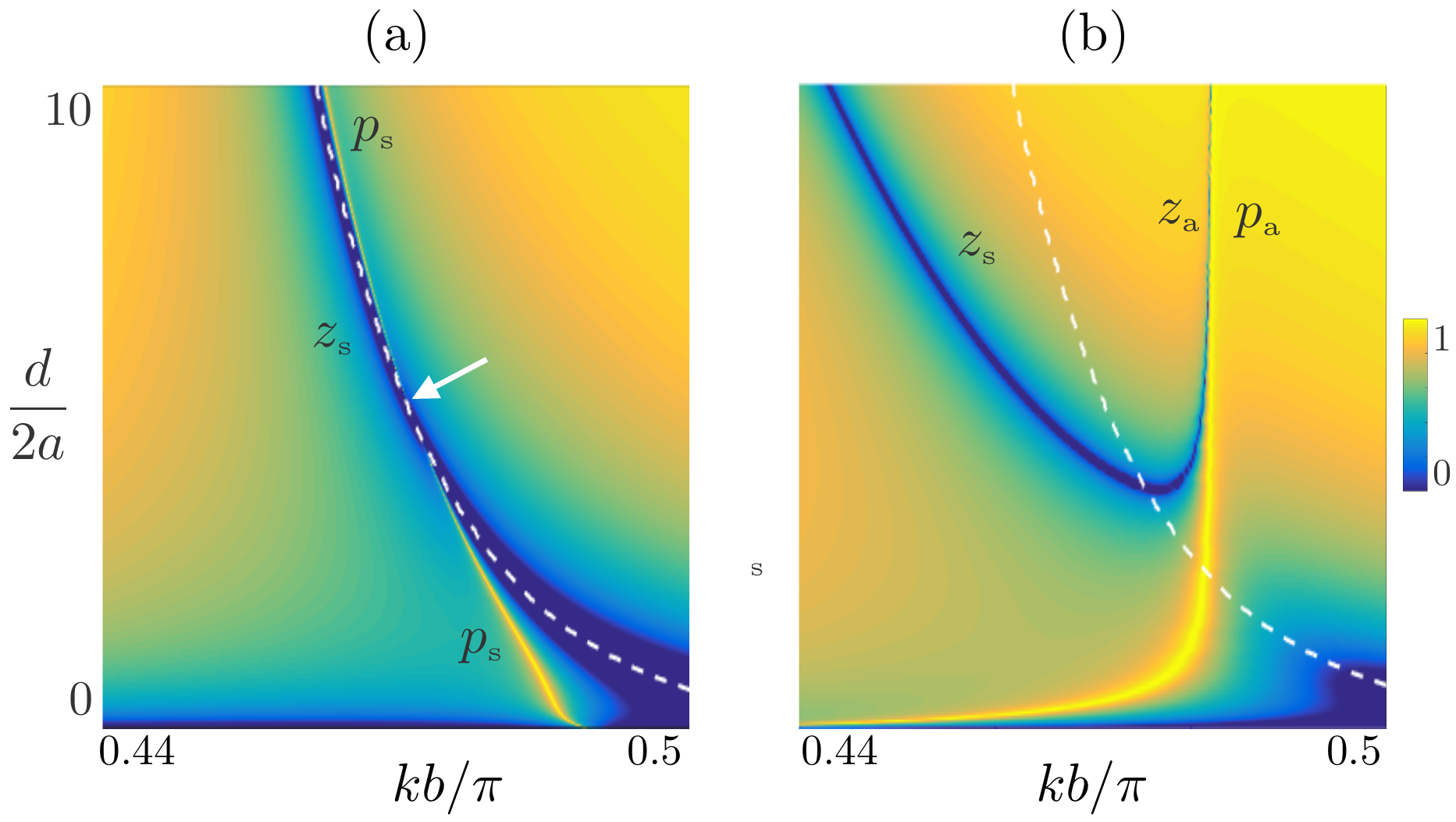}
	\caption{Transmission spectra (in log-scale) against the non-dimensional frequency $kb/\pi$ and the guide width $d/2a$. The dashed white lines show
	$k_0$ associated with zero-transmission for a single channel.
	  }
	\label{Fig14}
\end{figure}

\vspace{.3cm}
We end with a comparison between AIT and ATS and we report in figure \ref{Fig14} the transmission spectrum 
as the guide width $d/2a$ varies
  for $h=20a$ and for $h=a$. As a guideline, the dashed white line shows the trajectory of the zero-transmission for a single channel. 
For (a) $h=20a$, the zero-transmission follows the same path as for a single channel; it is 
 preceded or followed  by the perfect transmission resulting in a Fano resonance
	which disappears at the perfect tuning (white arrow). For (b)  $h=a$, 
	the two zero-transmissions result from the splitting of the single channel zero-transmission.

\section{Scattering by channels with monopolar and dipolar resonances }
\label{S5}
\begin{figure}[h!]
\centering
\includegraphics[width=.8\columnwidth]{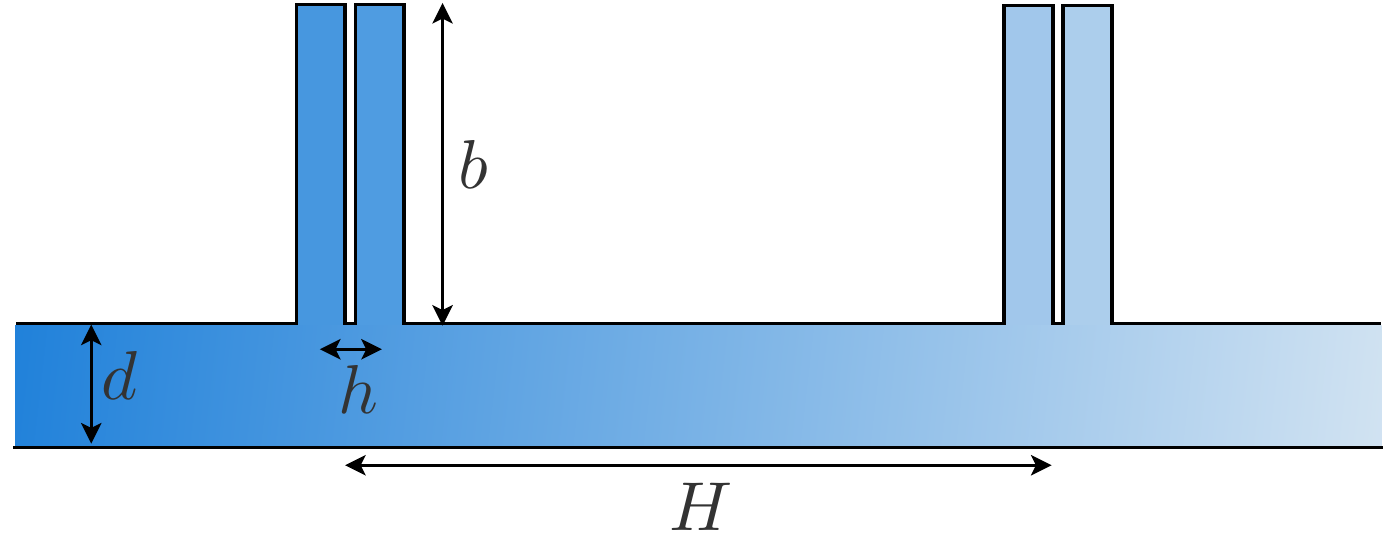}
	\caption{Arrangement of channels able to produce monopolar and dipolar resonances.}
	\label{Fig15}
\end{figure}

Owing to the previous analyses  for large or subwavelength separation distances, we now consider a combination of channels offering  monopolar and dipolar resonances. This is done using  two scatterers with spacing $H$ on the scale of the wavelength, each  scatterer being composed by two channels with subwavelength separation distance $h$ (figure \ref{Fig15}). 

\begin{figure}[h!]
\centering
\includegraphics[width=1\columnwidth]{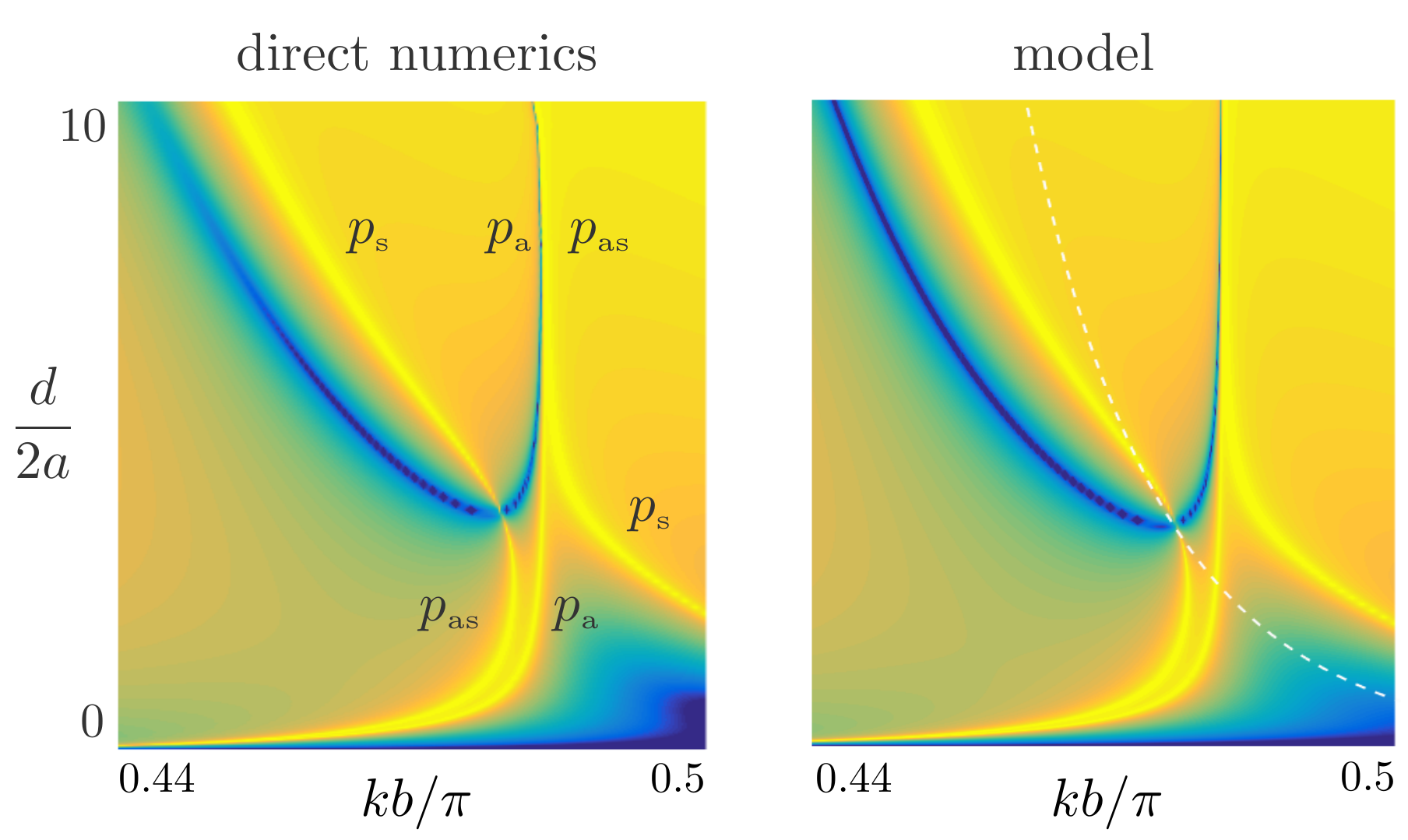}
	\caption{Transmission spectrum  for $H/2a=20$, $h/2a=1$ and  $b/2a=10$, from direct numerics and from \eqref{RTmixte}.  }
	\label{Fig16}
\end{figure}

\begin{figure}[h!]
\centering
\includegraphics[width=1\columnwidth]{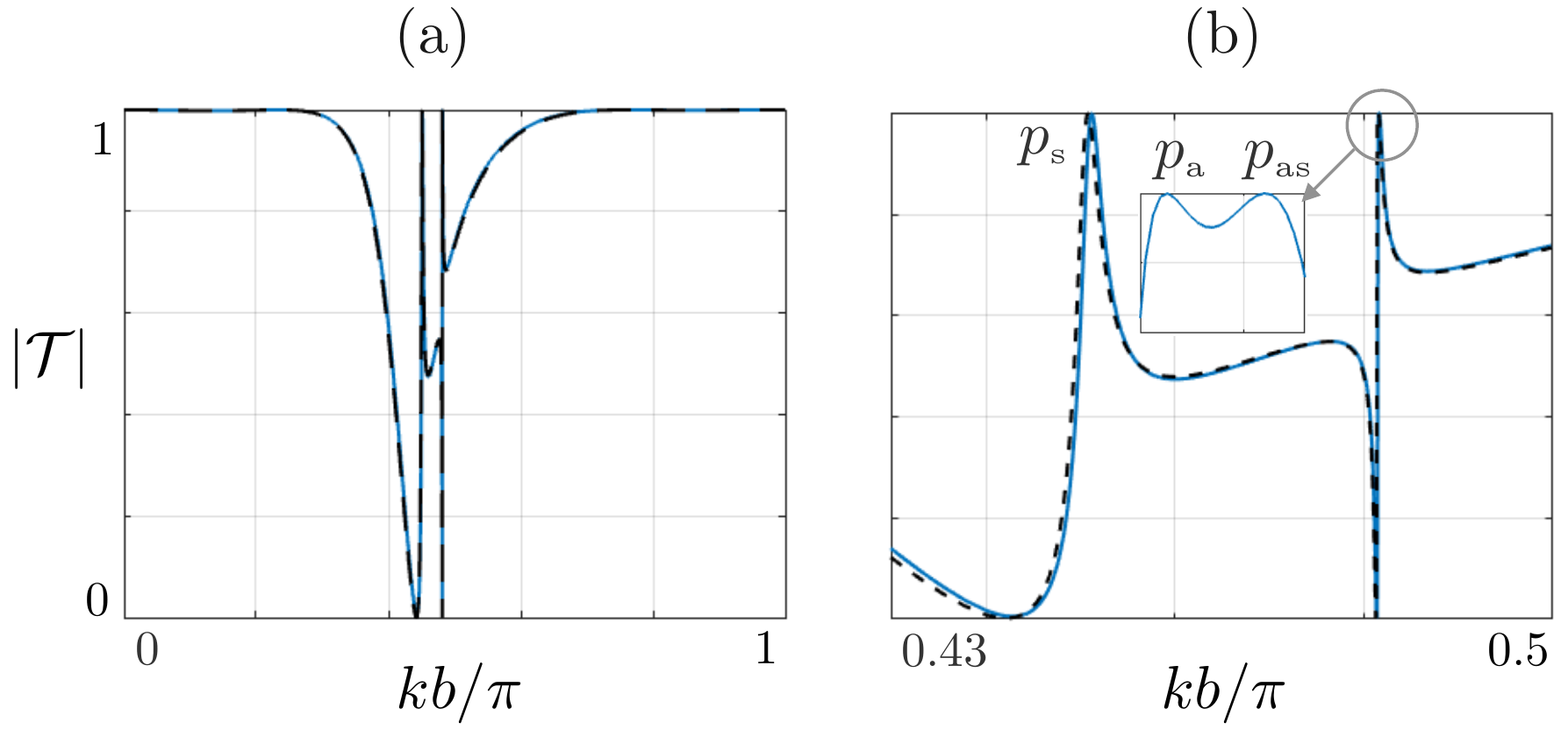}
	\caption{Transmission profile for $d/2a=10$ from direct numerics (plain line) and from \eqref{RTmixte} (dashed black line), (b) shows a magnification on the two peaks and the inset a magnification on the second transmission peak $kb/\pi \in 0.48170+(0, 8.10^{-5})$.  }
	\label{Fig17}
\end{figure}

\vspace{.1cm}
The  corresponding one-dimensional  model is straightforward and it corresponds to a mix of the models a and b. Indeed, we have to consider two junctions with separation distance $H$ as in the model a. Next at each junction \eqref{saut2} applies for a scatterer being composed by a pair of channels. Hence 
the solution reads as in \eqref{solfar} replacing $h$ by $H$ and $(\als,\ala)$ by $(\bes,\bea)$. As a result, the scattering coefficients $({\cal R},{\cal T})$ read

\beq\label{RTmixte}
\toutin
\dsp {\cal R}=\frac{1}{2}\left(\frac{\cZs}{\cZs^*}-\frac{\cZa}{\cZa^*}\right),\quad 
\dsp {\cal T}=\frac{1}{2}\left(\frac{\cZs}{\cZs^*}+\frac{\cZa}{\cZa^*}\right),\vspace{.4cm}\\
\dsp \cZs=(1+i\ta)\left(1+\frac{\bea\bes}{4}\right)-i\frac{\bes}{kd}-\bea kd \ta,\vspace{.3cm}\\
\dsp \cZa=(1+i\ta)\left(1+\frac{\bea\bes}{4}\right)+\frac{\bes}{kd}\ta+i\bea kd, 
\toutout\eeq
where $\ta=\tan (kH/2)$.
Figure \ref{Fig16} shows the transmission spectrum by the four channels with $H/2a=20$ and $h/2a=1$  (same representation as in figure \ref{Fig14}). We notice  the good overall agreement  between the direct numerics  and the model   \eqref{RTmixte}  and this is further illustrated in figure \ref{Fig17} with the profile of $|{\cal T}|$ for $d/2a=10$.


\vspace{.3cm}

The overall appearance of the spectrum in figure \ref{Fig16} is very similar to that of a single pair of channels reported in figure  \ref{Fig14}(b) which suggests that the dipolar resonance is more robust than the monopolar one. The branch  $\pa$ is reco\-ve\-red almost unchanged but it is now sandwiched  bet\-ween two   branches being the branch $\ps$ and a new branch that we term $\pas$. 
 \begin{figure}[b!]
\centering
\includegraphics[width=1\columnwidth]{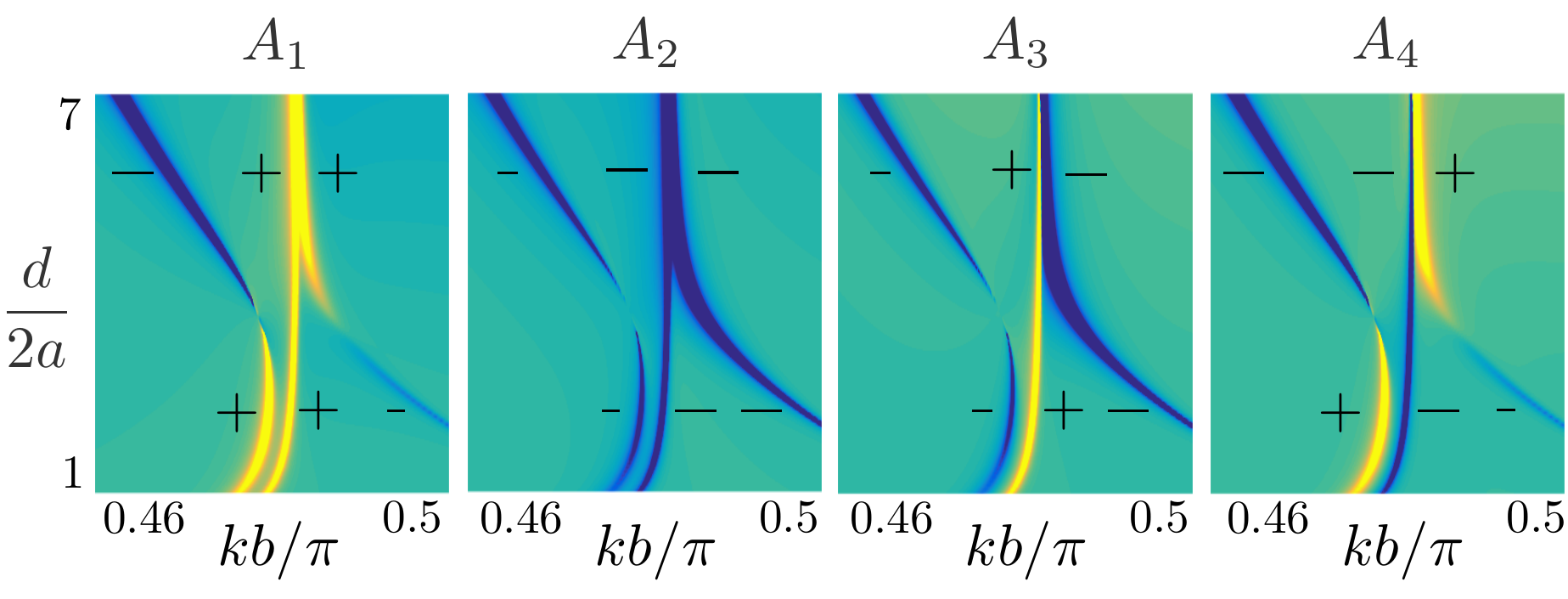}
	\caption{Real parts of the amplitudes $A_n$, $n=1,\cdots 4$, at the tops of the 4 channels.}
	\label{Fig18}
\end{figure}
To  identify the different branches, we considered the real parts of the amplitudes $A_n$, $n=1,\cdots,4$ against non-dimensional frequency and guide width (figure \ref{Fig18}). On the central  branch, $(A_1,A_2,A_3,A_4)=(+,-,+,-)$ which means that each pair of channel realizes  the perfect-transmission $\pa$ as if it was alone; this is why we continue to name it $\pa$.  Surrounding this central branch, the side branches experience a kind of avoiding crossing at $d/2a\sim 3.5$. We continue to term $\ps$ the parts of these branches corresponding to the four channels with in-phase oscillations that is $(A_1,A_2,A_3,A_4)=(-,-,-,-)$. Indeed, along $\ps$ the two  pairs of channels behave as two  single channels of width $4a$ with separation distance $H$. Eventually, $\pas$ corresponds to   $(A_1,A_2,A_3,A_4)=(\pm,\mp,\mp,\pm)$; this is a new resonance with out-of-phase oscillations of the contiguous  channels within a pair, and out-of-phase oscillations of the two pairs. 
From figure \ref{Fig18}, the parts of the side-branches associated with $\ps$ and $\pas$ have been identified (see figure  \ref{Fig16}) and we report 
 in  figure \ref{Fig19} the fields at the three perfect-transmissions for $d/2a=3$ and 5. Both  $\ps$ and $\pas$ are  associated with a Fano resonance and, at the perfect tuning, which occurs at $d/2a\simeq 3.5$, they exchange  places. Additional results are given in  appendix \ref{app3}.  

\begin{figure}[h!]
\centering
\includegraphics[width=1\columnwidth]{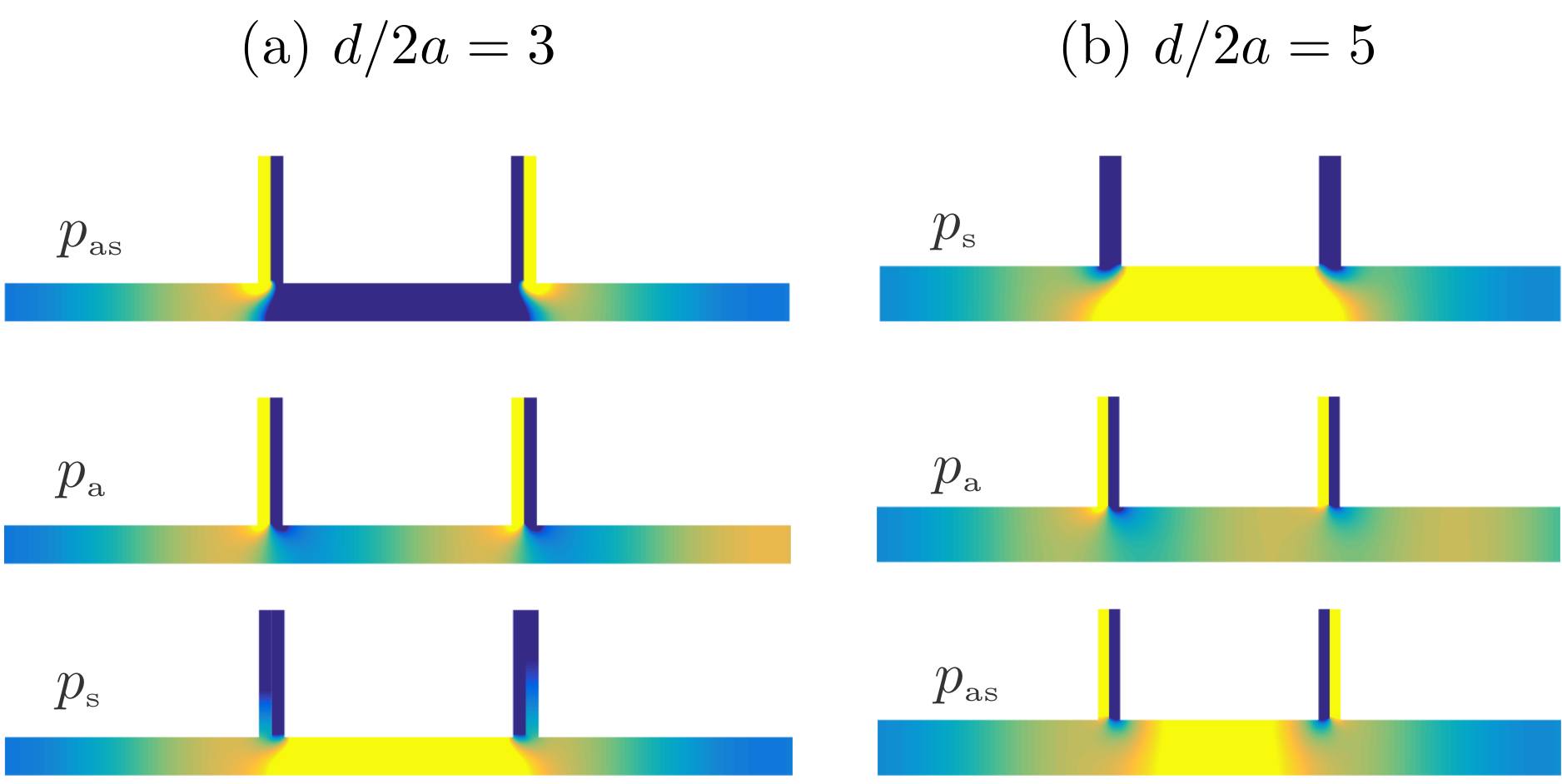}
	\caption{Real parts of the fields at  perfect-transmissions $\pas$, $\pa$ and $\ps$; from top to bottom, the frequency is increasing.   }
	\label{Fig19}
\end{figure}

\vspace{.3cm}

\begin{figure}[b!]
\centering
\includegraphics[width=1\columnwidth]{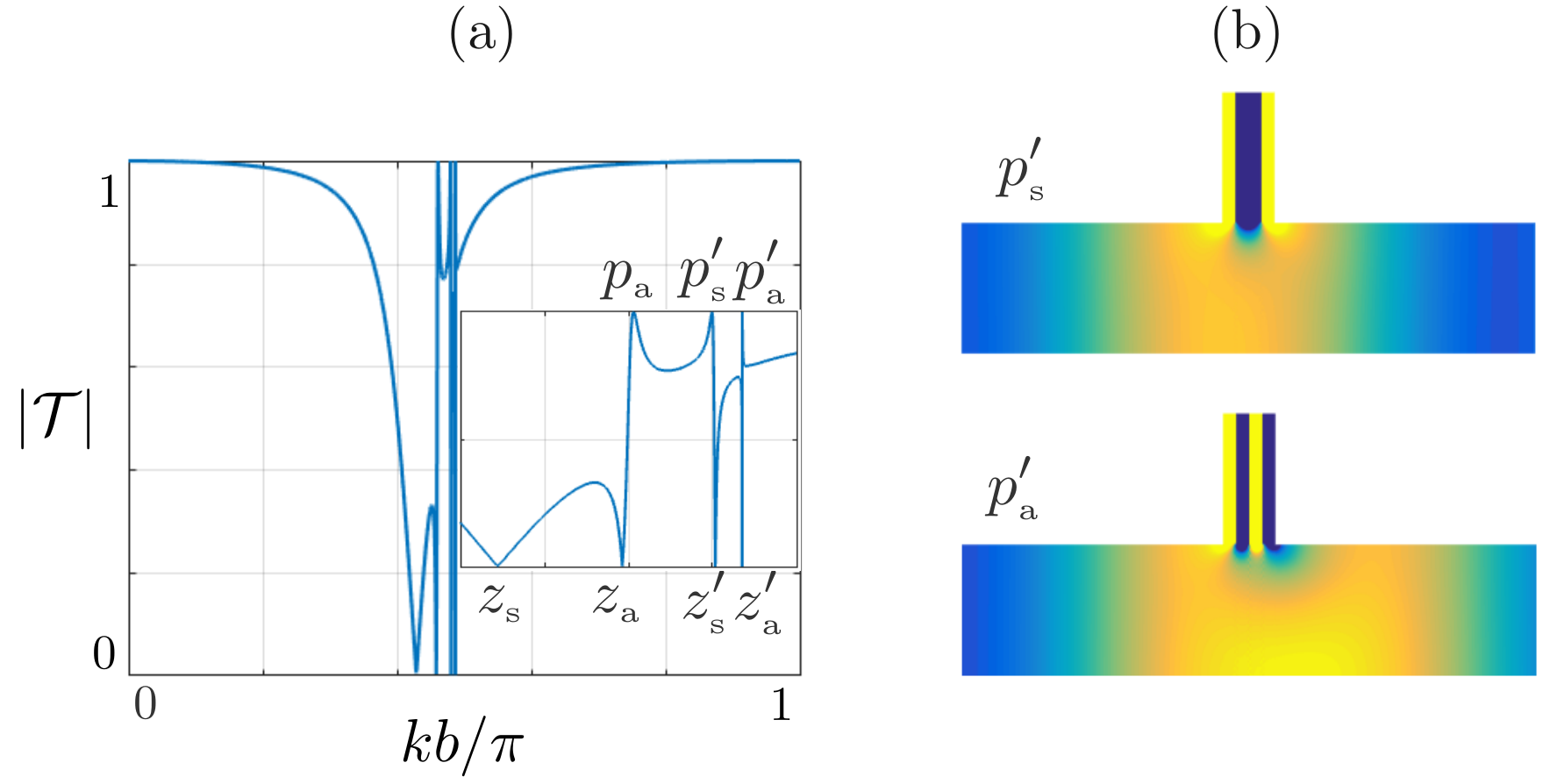}
	\caption{(a) Transmission $|{\cal T}|$ against non dimensional frequency for four contiguous  channels; the inset shows the range $kb/2\pi\in (0.42,0.5)$. (b) Real parts of the fields at the perfect transmissions $(\ps',\pa')$ corresponding to the new resonances of the set of four channels. }
	\label{Fig20}
\end{figure}

We conclude our study with a  remark which is evident in light of 
what has gone before. 
The model above which mixes models a and b and provides \eqref{RTmixte} is valid only when $H$ is on the scale of the wavelength.
  If $H$ becomes subwavelength, the  branch $\pas$ cannot exist since it involves interference between the two pairs of channels. Instead, new branches appear which are associated with  resonances of a single scatterer comprised  of four contiguous channels. These new arrangements are shown in figure \ref{Fig20}. 
We recover $\zs$ when the four channels behave as a single channel of width $8a$ and $(\za,\pa)$ when the  first two and the last  two channels behave as a pair of channels of width $4a$. Next,  the first new resonance is symmetric with 
the two central channels and the side channels oscillating  out-of-phase. The second new resonance is antisymmetric with the four successive channels oscillating out-of-phase. Each new resonance produces  an additional  zero- and perfect-transmission $(\zs',\ps')$ and $(\za',\pa')$.  
%

\vspace{1cm}

To account for these new resonances, we should apply the analysis conducted for a set  of two channels (model b) to the set of four channels. Such analysis  would involve five elementary problems instead of the three sketched in figure \ref{Fig5} (more generally, a set of $N$ channels with subwavelength extend would involve $(N+1)$ elementary problems).    
Note that the accumulation of resonances for large $N$ while keeping constant the small extend $2Na$ has been considered in \cite{jan2018}   (see figure 2 in this reference and figure 3 in the associated supplementary report). The succession of transmission profiles in figures \ref{Fig6} ($N=1$),  \ref{Fig11} ($N=2$) and \ref{Fig20} ($N=4$) show how this  limit would be reached.

 \section{Conclusion}
We have analyzed the scattering of waves in a guide attached to two side channels  with quarter-wavelength resonances using the method of matched asymptotic expansions. The asymptotic procedure includes the analysis of a problem of potential flow 
in a region local to the side-channel junction. 
Solutions are naturally sensitive to the local geometry at the junction especially whether it contains one or a pair of side channels.
It results two different models 
which are increasingly accurate in the limit of large and small channel separation.

The present analysis applies to analog systems where ATS and EIT have been discussed, including acoustic waves in a guide loaded by Helmholtz resonators \cite{cheng2019}, elastic guided waves such as Lamb waves in a plate or  Love waves in a  layer supporting pillars \cite{jin2018,liu2019}. 
As previously said, interesting extensions arise in the case of multiple channels with small separation distances as considered in \cite{porter2018,jan2018,cervenka2019}. In the case where the set of channels still has a subwavelength extension the present analysis is applicable and it will provide  jump conditions applying at a single junction. 
In the case where the 
array of channels has an extent of 
 the order of greater than the wavelength,  homogenization of periodic medium can be used which would provide  an effective guided wave propagation as done in \cite{maurel2019} in the context of water waves. 
 Eventually, we note a recent closely related study by \cite{schnitzer2022}  revealing strong dipolar resonances of a channel partially partitioned by a thin wall and connected to a half space.
This resonance is associated with a quasi-trapped mode whose origin is the perfect trapped-mode of the semi-infinite  channel (partially partitioned). 
It would be interesting to understand if this resonance can be used and combined to our totally partitioned  channels.

\appendix
\section{Explicit form of the parameters $(\cA,\cAs)$ in the model a}
\label{app}
In this appendix, we provide explicit solutions of  the elementary problems set on $(\phineu,\Phis)$ entering in the  \eqref{psi}. The elementary solutions $(\phineu,\Phis)$  satisfy the Laplace equation and  their asymptotic behaviours are given by  \eqref{ag1}.
We rescale variables on the  opening width $2a$ with 
  $$x=aX, \qquad y+d=aY,$$  
and we let $Z = X+ \ci Y$.  The boundary conditions are $\partial_X\Phi=0$ on $\{X=\pm 1,Y>0\}$, $\partial_Y\Phi=0$ on $\{X\in(-\infty,\infty),Y=0\}$ and on $\{X\in(1,\infty),Y=\frac{1}{\th}\}$. 
To find the solution, we use complex analysis and  
 introduce the
Schwarz-Christoffel mapping defined by
\begin{equation}
 \frac{dZ}{d\zeta} = - \frac{2 \sqrt{1-\zeta^2}}{\pi (\zeta^2-\sigma^2)}
 \label{eq:4.18}
\end{equation}
($\sigma > 0$) which integrates to 
\begin{equation}
 Z = \frac{2}{\pi} 
 \sin^{-1}\zeta + \frac{\sqrt{1-\sigma^2}}{\pi \sigma} 
 \ln \left( \frac{\sigma \sqrt{1-\zeta^2} + \zeta \sqrt{1-\sigma^2}}
 {\sigma \sqrt{1-\zeta^2}-\zeta \sqrt{1-\sigma^2}}
 \right).
 \label{eq:4.19}
\end{equation}

\vspace{.4cm}

\begin{figure}[h!]
\centering
\includegraphics[width=1\columnwidth]{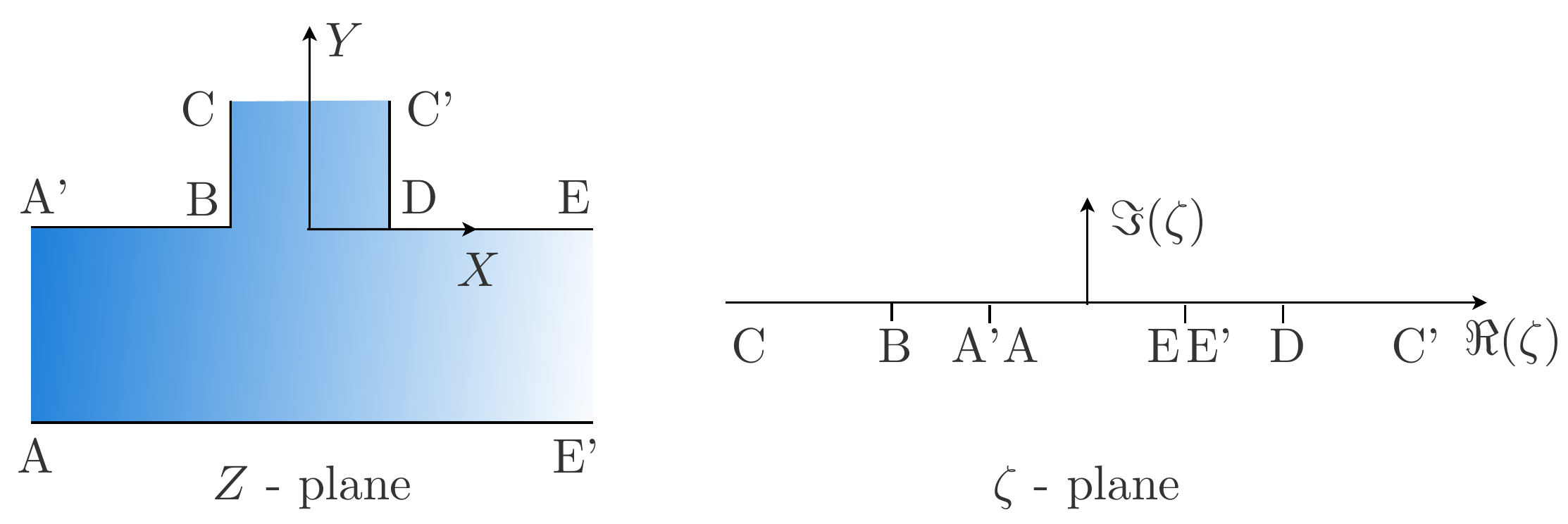}
	\caption{Conformal mapping  of the junction between the guide and the side channel.}
	\label{Fig21}
\end{figure}

The result of the mapping is shown in figure \ref{Fig21}.
The lower wall $\{X\in(-\infty,\infty),Y=0\}$ is mapped to 
$\Re\{ \zeta \} \in (-\sigma,\sigma)$; the right part of the upper wall 
$\{ X \in(1,\infty),Y=\frac{1}{\th}\}$ is mapped to $\Re \{ \zeta \} \in(\sigma,1)$; the vertical wall of the side channel  
$\{X=1,Y \in (\frac{1}{\th},\infty)\}$ is mapped to $\Re \{ \zeta \} \in(1, \infty)$;
and mappings of boundaries in $X < 0$ to the negative real-$\zeta$ axis 
apply in a similar manner.
We note the relationships 
$$\frac{\sqrt{1-\sigma^2}}{\sigma} = \frac{1}{\th},\quad
\sigma = \frac{\th}{\sqrt{1+\th^2}}.$$

\vspace{.4cm}
Harmonic potentials satisfying the homogeneous Neumann condition on the 
real $\zeta$-axis can be written as a combination of 
\begin{equation}
\frac{1}{2\pi}\ln\left(\frac{ |\zeta-\sigma|}{2\sigma}\right)
\quad\text{and}\quad  \frac{1}{2\pi} \ln\left(\frac{|\zeta+\sigma|}{2\sigma}\right),
 \label{eq:4.20}
\end{equation}
up to a constant. 
\vspace{.3cm}

We inspect the behaviours of the potentials when $X\to +\infty$ ($|\zeta-\sigma| \sim 0$), $X\to -\infty$ ($|\zeta+\sigma| \sim 0$)  in the main channel and when $Y\to+\infty$ ($|\zeta|\to \infty$) in the side channel.

\noindent $\bullet$ When $X\to+\infty$,
 we determine from (\ref{eq:4.19}) that
 \beq\begin{array}{l}Z \sim \frac{2}{\pi} \sin^{-1} \sigma + \frac{\sqrt{1-\sigma^2}}
 {\pi \sigma} \ln \left( \frac{2 \sigma (1-\sigma^2)}{-(\zeta-\sigma)} \right),
 \end{array}\eeq 
and consequently
  \beq\begin{array}{l}X \sim \frac{2}{\pi} \sin^{-1} \sigma + \frac{\sqrt{1-\sigma^2}}
 {\pi \sigma} \left( \ln (1-\sigma^2)-\ln \left(\frac{|\zeta-\sigma|}{2\sigma}\right) \right),
 \end{array}\eeq 
(which   tends to $+\infty$).

\noindent $\bullet$
Similarly,  as $X\to-\infty$ 
we have
  \beq\begin{array}{l}X \sim -\frac{2}{\pi} \sin^{-1} \sigma - \frac{\sqrt{1-\sigma^2}}
 {\pi \sigma} \left( \ln (1-\sigma^2)-\ln \left(\frac{|\zeta+\sigma|}{/2\sigma}\right) \right).
 \end{array}\eeq 

\noindent $\bullet$ Finally, in the limit
$|\zeta| \to +\infty$, we find  that
  \beq\begin{array}{l}Z \sim \frac{2 \ci}{\pi} \ln (-2\ci \zeta) + \frac{2 \ci }{\pi}
 \frac{\sqrt{1-\sigma^2}}{\sigma} \tan^{-1} \left(\frac{\sqrt{1-\sigma^2}}{\sigma}\right),
 \end{array}\eeq 
meaning that
 \beq\begin{array}{l}
 Y \sim \frac{2}{\pi} \ln (2 |\zeta|) + \frac{2}{\pi}
 \frac{\sqrt{1-\sigma^2}}{\sigma} \tan^{-1} \left(\frac{\sqrt{1-\sigma^2}}{\sigma}\right),
 \end{array}\eeq 
 (which tends to $+\infty$).

We deduce in terms of original variables that 
\beq
\toutind
x\to +\infty, & \frac{1}{\pi}\ln\frac{|\zeta-\sigma|}{2\sigma}\sim -\frac{x}{d}-\frac{\cA}{2},\vspace{.2cm}\\
x\to -\infty, &\frac{1}{\pi}\ln\frac{|\zeta+\sigma|}{2\sigma}\sim \frac{x}{d}-\frac{\cA}{2},\vspace{.2cm}\\
y\to+\infty, & \frac{2}{\pi}\ln |2\zeta|\sim \frac{y}{a}+\frac{2}{\pi\th}\tan^{-1}\th
\toutout
\eeq
where $\cA/2=-\frac{2\th}{\pi}\tan^{-1}\th+\frac{1}{\pi}\ln(1+\th^2)$ (we have used that $\tan^{-1}(1/\th)+\tan^{-1}\th=\pi/2$). According to the above behaviour, we recognize $(\phineu,\Phis)$ of the form
\beq
\phineu=\frac{1}{\pi}\ln\frac{|\zeta+\sigma|}{|\zeta-\sigma|},\quad 
\Phis=\frac{1}{\pi}\ln\frac{|\zeta+\sigma||\zeta-\sigma|}{4\sigma^2}+\frac{\cA}{2}.
\eeq

Indeed, we have
\beq\begin{array}{l}
\phineu\underset{x\to+\infty}{\sim}-\frac{1}{\pi}\ln\frac{|\zeta-\sigma|}{2\sigma}\sim 
\frac{x}{d}+\frac{\cA}{2},\vspace{.2cm}\\
\phineu\underset{x\to-\infty}{\sim}\frac{1}{\pi}\ln\frac{|\zeta+\sigma|}{2\sigma}\sim 
\frac{x}{d}-\frac{\cA}{2},\vspace{.2cm}\\
\phineu\underset{y\to+\infty}{\sim}0,
\end{array}\eeq
(as in \eqref{ag1}) which provides $\cA$. We also have  

\beq\begin{array}{l}
\Phis\underset{x\to+\infty}{\sim}\frac{1}{\pi}\ln\frac{|\zeta-\sigma|}{2\sigma}+\frac{\cA}{2}
\sim 
-\frac{x}{d},\vspace{.2cm}\\
\Phis\underset{x\to-\infty}{\sim}\frac{1}{\pi}\ln\frac{|\zeta+\sigma|}{2\sigma}+\frac{\cA}{2}\sim 
\frac{x}{d},\vspace{.2cm}\\
\Phis\underset{y\to+\infty}{\sim}\frac{2}{\pi}\ln\frac{|\zeta|}{2\sigma}+\frac{\cA}{2}
\sim \frac{y}{a}+\cAs,
\end{array}\eeq
with $\cAs=\frac{2}{\pi}\left(\ln \frac{1+\th^2}{4\th}+(1/\th-\th)\tan^{-1}\th\right)$
(as in \eqref{ag1}) which provides $\cAs$.

\section{The effective parameters}
\label{app2}

\begin{figure}[h!]
\centering
\includegraphics[width=1\columnwidth]{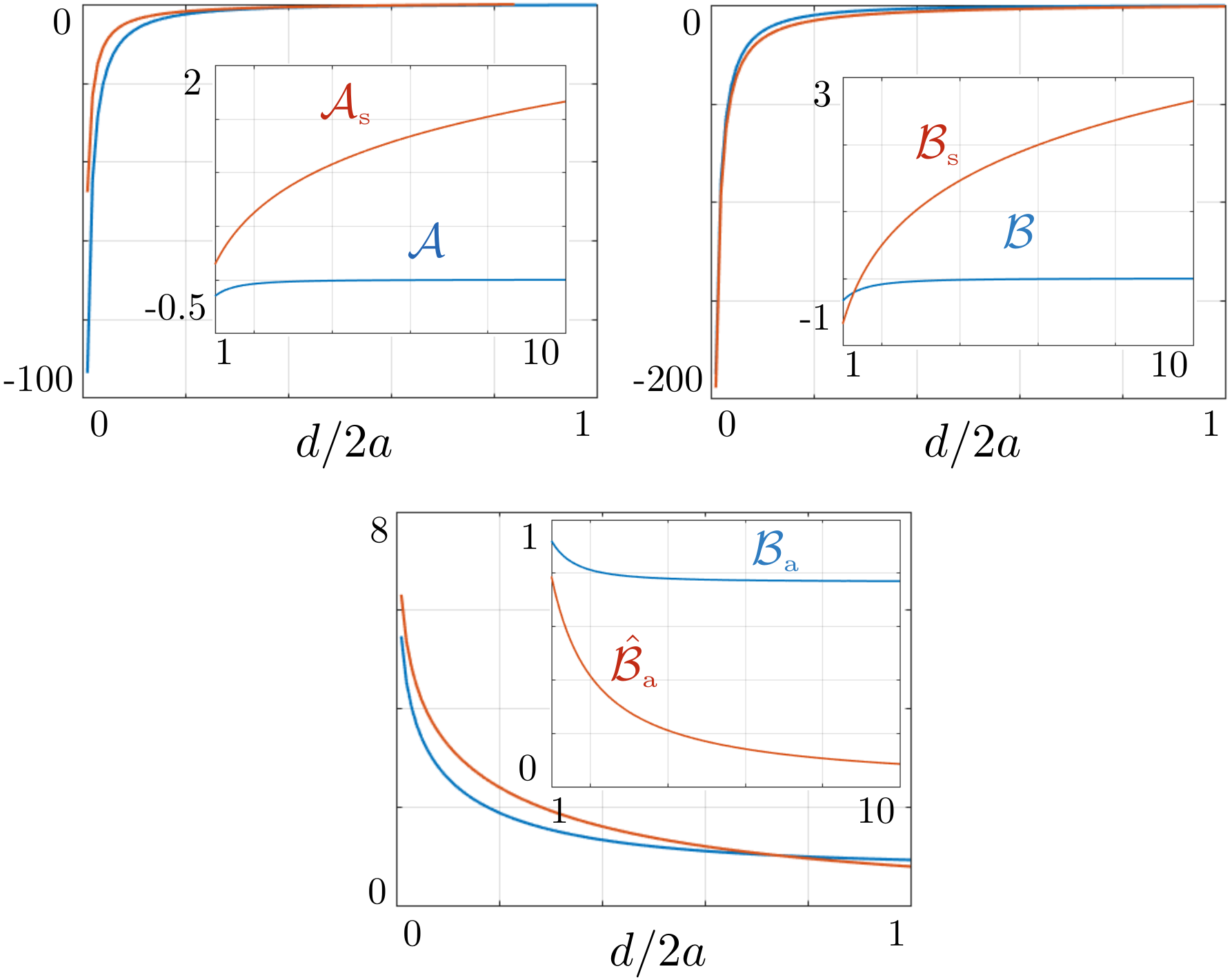}
	\caption{ Variations of $(\cA,\cAs)$ against $d/2a$ (from \eqref{defAAs}), and of $(\cB,\cBs,\cBa,\cBac)$  (for $h=a$, computed numerically).}
	\label{Fig22}
\end{figure}

\begin{figure}[h!]
\centering
\includegraphics[width=1\columnwidth]{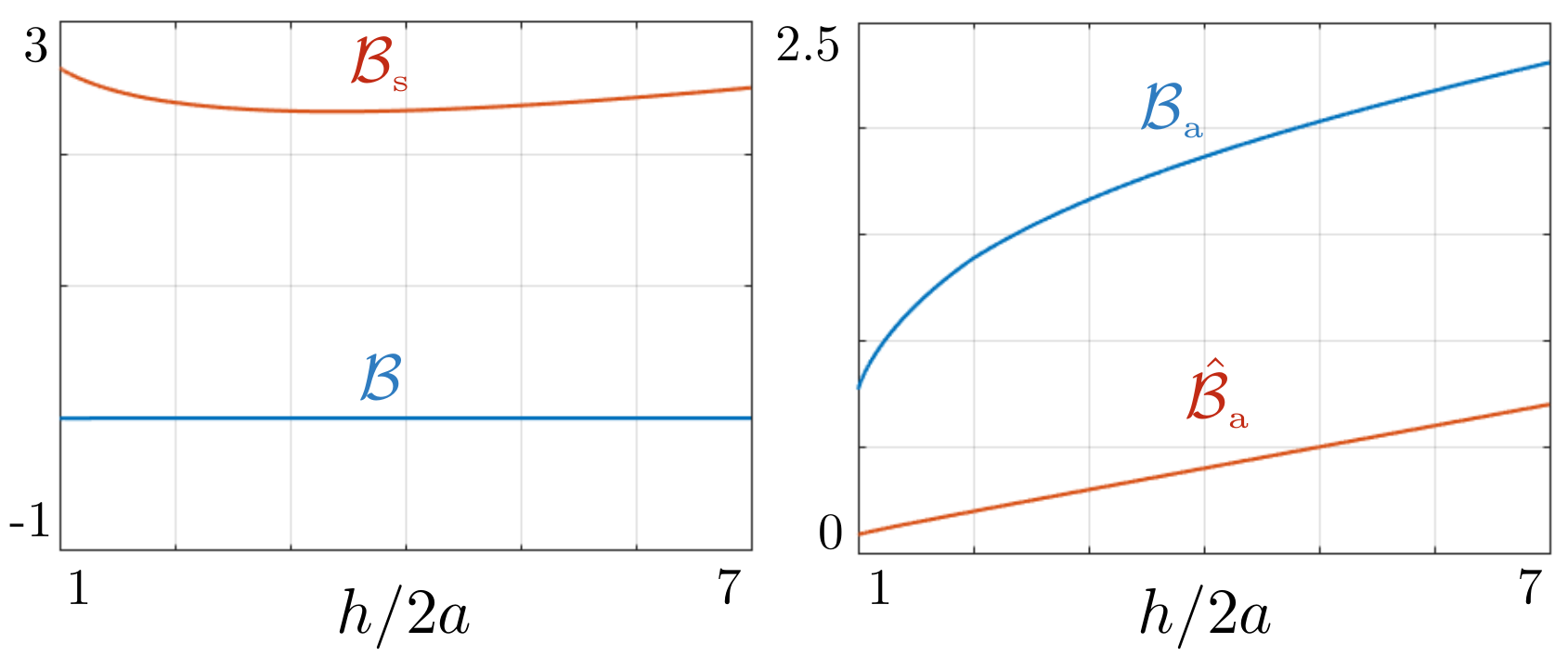}
	\caption{ Variations of $(\cB,\cBs,\cBa,\cBac)$  against $h/2a$ (for $d/2a=10$).}
	\label{Fig23}
\end{figure}

The variations of $(\cA,\cAs)$ for a single channel and of $(\cB,\cBs,\cBa,\cBac)$ for a pair of channels against $d/2a$ are shown in figure \ref{Fig22}; for a pair of the channels, the variations of the effective parameters against $h/2a$ are shown in  \ref{Fig23}.
We note that the parameters $(\cA,\cAs)$ and $(\cB,\cBs)$ have strong variations when $d/2a<\sim 0.2$ responsible for the shifts of the frequency realizing perfect transmission in figures \ref{Fig14} (shift to higher frequencies for a single channel and shift to a lower frequency for a pair of channels).
 
In \eqref{K0} and \eqref{gg1}, we provide an approximate expression  of $T$ around $kb=\pi/2$. Orginally, the Taylor expansion provides \eqref{Tappb} with 
\beq\nonumber 
k_0^\pm=\frac{\pi}{2b}\left(1-\frac{a}{2b}\left(\cBs-\cB+\cBa\pm\sqrt{\Delta}\right)\right),
\eeq
with $\Delta=(\cBs-\cB-\cBa)^2-4 \cBac^2$ and
\beq\nonumber\toutin
 k^+=\frac{\pi}{2b}\left(1-\frac{a}{b}\cBa-i\frac{\pi ad}{4b^2} \cBac^2\right),
\vspace{.2cm}\\
 k^-=\frac{\pi}{2b}\left(1-\frac{a}{b}\cBs\right)-i\frac{2\th}{b}.
\toutout
\eeq
From figures \ref{Fig22} and for  $d/2a>\sim 2$, we notice  that $\cAs\gg \cA$, hence $\ksz b\sim \pi/2(1-\frac{a}{b}\cAs)$. We also notice that  $\cBs\gg \cB$ and $(\cBa+\cBs)\simeq 2\cAs$. Hence, defining $2\delta_r=(\cBs-\cBa)$ (hence $\cBs=\cAs+\delta_r$ and $\cBa=\cAs-\delta_r$),  we obtain the form of $\ksz^\pm$ in \eqref{K0} and $k^\pm$ in \eqref{gg1}.

\section{Transmission by four channels}
\label{app3}

\begin{figure}[h!]
\centering
\includegraphics[width=1\columnwidth]{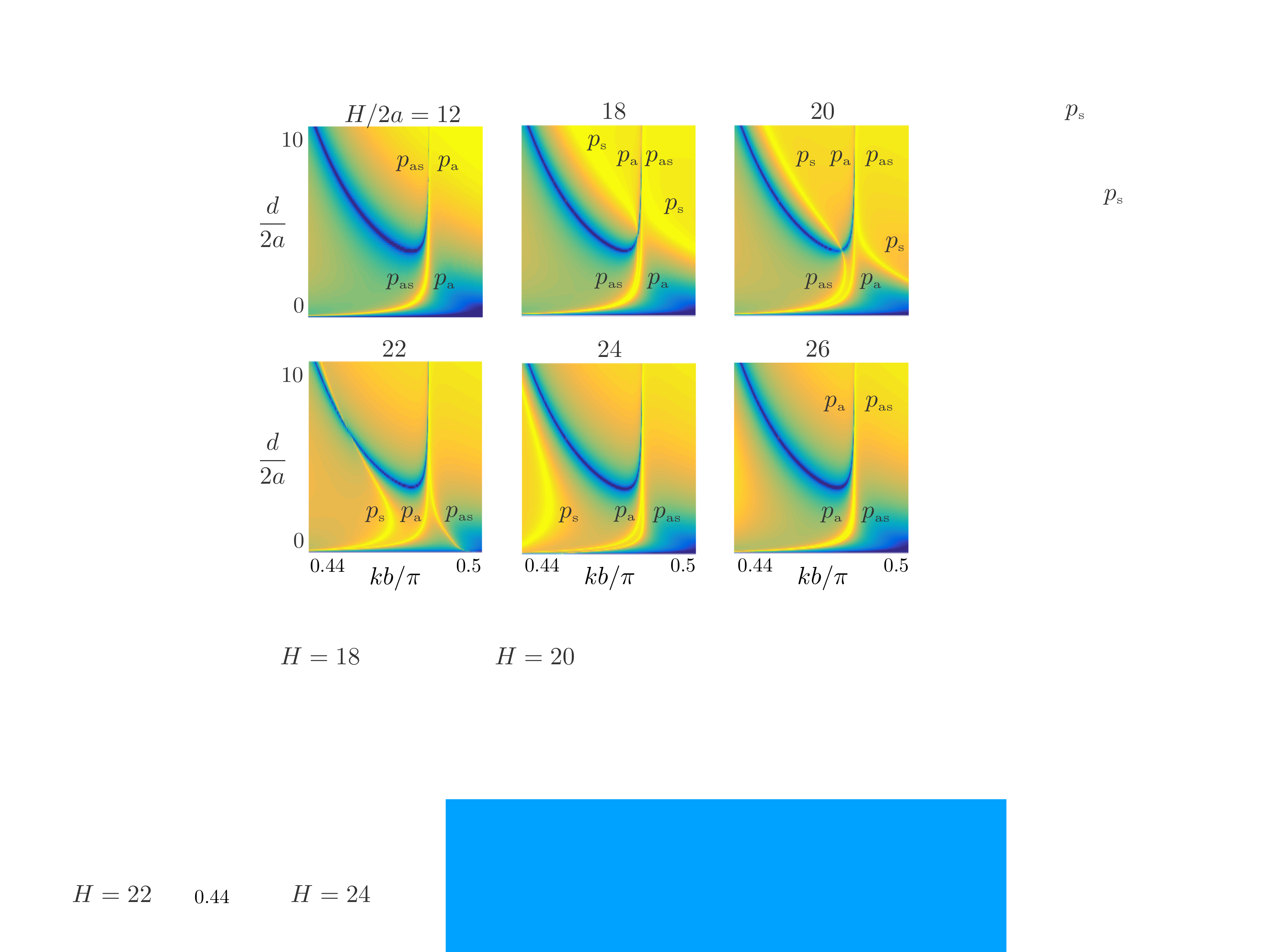}
	\caption{ Transmission spectra (in log-scale) against non-dimensional frequency $kb/\pi$ and non-dimensional guide width $d/2a$ for different spacings $H/2a$.}
	\label{Fig26}
\end{figure}

We provide in figure \ref{Fig26}  additional results on the transmission spectra for different values of $H/2a$ (same representation as in figure \ref{Fig16}). The different branches have been identified thanks to the amplitudes in the four channels as in figure \ref{Fig18}.  

We have already noticed that  the central branch corresponds to perfect transmission by  each pair of channels which behaves as if it was alone. Accordingly, this branch is almost identical to that reported in figure \ref{Fig14} for a single pair of channels, and we now notice that this remains true whatever the value of $H$. Next, the branches $\pa$ and $\pas$  remain close to each other whatever the value of $H$ but their relative positions change (see $H/2a=12$ and 26).
This rearrangement is attributable to the interaction of $\pas$ with the branch $\ps$  for $H/2a$ between roughly 16 and 21. This later  arrives from the high frequency region ($0.5<kh/\pi<1$)  and it is shifted to lower frequency as $H$ increases, a  trajectory which was already observed for two channels, see figure \ref{Fig8}. 
It is worth noticing that the conversion of $\pas$ to $\ps$ on the lower branch correspond to a sudden change in the amplitudes  $A_1$ and $A_4$  when the Fano resonance disappears (see figure \ref{Fig18}). In contrast the conversion of $\ps$ to $\pas$ on the upper branch does not conduct to the suppression of the associated perfect-transmission. Instead, the amplitudes $A_1$ and $A_4$ vanish and the Fano resonance is supported by the two central channels only.

\end{document}